\documentclass[%
 reprint,
 amsmath,amssymb,
 aps,
floatfix,
]{revtex4-2}

\usepackage{subcaption}

\usepackage{graphicx}
\usepackage{dcolumn}
\usepackage{bm}
\usepackage{multirow} 
\usepackage{float} 
\restylefloat{table} 

\usepackage[table]{xcolor}
\definecolor{blue}{rgb}{0.36, 0.54, 0.85}
\definecolor{amaranth}{rgb}{0.9, 0.17, 0.31}
\definecolor{pink}{rgb}{0.87, 0.56, 0.81}
\definecolor{ao}{rgb}{0.0, 0.5, 0.0}
\definecolor{maroon}{rgb}{0.76, 0.13, 0.28}
\definecolor{cardinal}{rgb}{0.77, 0.12, 0.23}
\definecolor{lightcardinal}{rgb}{0.97, 0.42, 0.53}
\definecolor{frenchlila}{rgb}{0.53, 0.38, 0.56}
\definecolor{yellow}{rgb}{1.0, 1.0, 0.87}
\definecolor{lightseagreen}{rgb}{0.45, 0.85, 0.58}
\definecolor{gray}{rgb}{0.9, 0.9, 0.9}
\definecolor{lightblue}{rgb}{0.66, 0.84, 0.96}

\usepackage[pagebackref=false,
			hyperindex=true,
			citecolor=blue,
			colorlinks=true,
			linkcolor=blue,
			urlcolor=blue]{hyperref}

\newcommand{\param}{\ensuremath{\vec{\theta}} }
\newcommand{\paramS}{\ensuremath{\vec{\sigma}} }

\newcommand{\hyp}{\ensuremath{{\cal H}}}
\newcommand{\whtl}{\ensuremath{{\cal W}}}

\newcommand\abs[1]{\left|#1\right|}

\newcommand{\mrm}[1]{\mathrm{#1}}

\begin{document}

\preprint{APS/123-QED}
\title{Beyond Gaussian Assumptions: A new robust statistical framework for gravitational-wave data analysis}
\date{\today}
\author{Argyro Sasli$^{1,2,3}$, Minas Karamanis$^{4}$, Nikolaos Karnesis$^{3}$, Michael W. Coughlin$^{1,2}$, Vuk Mandic$^{1}$, Uro\v{s} Seljak$^{4}$ and Nikolaos Stergioulas$^{3}$\\
$^{1}$ School of Physics and Astronomy, University of Minnesota, 55455 MN, USA\\
$^{2}$ NSF Institute on Accelerated AI Algorithms for Data-Driven Discovery (A3D3),
MIT, Cambridge, MA 02139 and University of Minnesota, Minneapolis, MN 55455\\
$^{3}$ Department of Physics, Aristotle University of Thessaloniki, Thessaloniki 54124, Greece\\
$^{4}$ Physics Department, University of California and Lawrence Berkeley National Laboratory Berkeley, Berkeley, CA 94720, USA
}
\begin{abstract}
Many traditional algorithms applied in gravitational-wave astronomy rely on the assumption of Gaussian noise, a condition not always met. To meet this need, this study extends a robust statistical framework, advancing previous work on heavy-tailed likelihoods, that adapts the hyperbolic likelihood method for full frequency domain applications. The framework is designed to maintain high performance under ideal conditions while improving robustness against non-Gaussian noise and outliers in real-world data. We demonstrate the efficacy of this approach through two key case studies. The first case study analyzes a massive black hole binary merger in simulated Laser Interferometer Space Antenna (LISA) data with Gaussian noise, showing that the extended hyperbolic likelihood method performs comparably to the more commonly used Whittle likelihood. The second case study examines a stellar-mass black hole binary merger using real ground-based gravitational-wave data containing non-Gaussian noise or overlapping signals, where our framework exhibits increased robustness and yields more accurate parameter estimations. Our results show that the hyperbolic likelihood better captures the true noise distribution, providing a flexible and physically motivated alternative for GW data analysis across current and future detectors.
\end{abstract}

\maketitle

\section{Introduction}
The detection and analysis of gravitational waves (GWs) has opened a new era in astrophysics, allowing us to probe the universe in unprecedented ways. However, accurate estimation of parameters from GW sources is critically dependent on the quality of the data and the understanding of noise ~\cite{PhysRevD.107.063004, RevModPhys.94.025001, Littenberg:2014oda, Baghi2023qnq, Speri:2022kaq}. A common assumption in the parameter estimation (PE) process is that the noise in the detector data is Gaussian and stationary. This assumption simplifies the mathematical treatment, making the use of Gaussian likelihood functions a standard practice. However, this practice may introduce significant implications for the accuracy and reliability of the estimated parameters~\cite{Sasli:2023mxr, Sasli:2025tjf}, particularly due to the presence of noise and potential outliers in the observed signals.

During the period between November 2019 and March 2020, noise transients, commonly known as ``glitches"~\cite{LISAPathfinder:2022awx, Baghi:2021tfd}, were observed at an average rate of one every 0.32 minutes at the LIGO Hanford Observatory and one every 1.17 minutes at the LIGO Livingston Observatory. Moreover, 17 out of the 90 gravitational-wave events reported in the first three Gravitational Wave Transient Catalogs occurred within one second of a glitch \cite{GWTC-3}. Similar glitch rates have also been reported in the O4 run \cite{GWTC-4}. Importantly, subthreshold glitches (SNR $< 5$) can still overlap in time and frequency with astrophysical signals, subtly impacting parameter inference despite remaining undetected by conventional glitch-mitigation methods \cite{udall2025inferringspinsmergingblack}. Such unmitigated noise transients and other artifacts~\cite{Dey:2021dem,Baghi:2019eqo,PhysRevD.67.082003,Wu_2023,Johnson_2024} pose a significant challenge for both detection pipelines and downstream PE analyses~\cite{Abbott_2023, mitigating_glitches, macas, Zackay_2021}.

Traditional approaches, such as the widely used Gaussian and/or Whittle likelihood, while effective under ideal conditions, can be sensitive to departures from Gaussian noise. At the same time, noise in GW detectors is not perfectly stationary; its statistical properties can change over time~\cite{PhysRevD.102.084062}. The Gaussian likelihood assumption fails to capture these temporal variations, potentially leading to underestimates or uncertainties in the PE and a higher probability of systematic errors.

In practice, the data collected by detectors like Advanced LIGO~\cite{LIGO}, Advanced Virgo~\cite{Virgo} and KAGRA~\cite{KAGRA} often contain various artifacts and non-stationarities, imposing variations in the statistical properties of the noise over time.  With the better sensitivities of upcoming detectors, such as ground-based detectors like Einstein Telescope (ET)~\cite{ET}, Cosmic Explorer (CE)~\cite{CE}, NEMO~\cite{NEMO} and future spaced-based detectors like LISA~\cite{Amaro2017}, TianQin and Taiji~\cite{chinesedetectors}, signals will overlap in time
and frequency, resulting in noise uncertainties~\cite{articleWang}. In the case of space-based detectors, calibrating the relevant noise models is not entirely feasible~\cite{articleWang} (we refer the reader to ~\cite{articleWang} and references therein for more details in different data challenges in the case of space-borne detectors), as is done for ground-based detectors~\cite{Abbott_2020_det}.

Several analyses have been conducted with the aim of tackling the aforementioned issues~\cite{Cornish:2014kda,PRD_RJMCMC_1,PRD_RJMCMC_2,PRD_RJMCMC_3, Baghi_2023, Edwards:2020tlp, PhysRevD.92.064011, Martellini:2014xia, Hamimeche:2008ai,Verde_2003, Flauger:2020qyi,PhysRevD.60.021101, PhysRevD.65.122002, Legin_2023}. Here, we adopt the statistical framework that was previously introduced in~\cite{Sasli:2023mxr, heavy_stochastic} and extend it to the whole frequency domain for PE purposes.

In \cite{Sasli:2023mxr}, we introduced a heavy-tailed likelihood, more specifically the Hyperbolic likelihood $\Lambda_{\cal H}$~\cite{Eberlein2002TheGH, Prause1999TheGH, Sasli:2023mxr, Sasli:2025tjf}, derived from the Generalized Hyperbolic distribution~\cite{Prause1999TheGH}. This likelihood offers a robust framework for data analysis, effectively handling data outliers, noise non-stationarities, and potential inaccuracies in the noise Power Spectral Density (PSD) model. We applied this approach to examples from GW astronomy, specifically using synthetic datasets from the adopted LISA mission. This method has been demonstrated to yield more robust results in PE~\cite{Sasli:2023mxr, Sasli:2025tjf}, particularly in scenarios where the data deviated from the theoretical estimation of the PSD of the noise. We concluded that the Gaussian likelihood assumption typically underestimates the tails of the noise distribution~\cite{Sasli:2023mxr}. This can result in overly optimistic confidence intervals~\cite{Sasli:2023mxr}, where the true parameter values lie outside the estimated ranges more frequently than predicted. This misinterpretation can hinder the reliability of scientific conclusions drawn from GW observations. However, we limited the application to a small frequency segment and did not expand the analysis to the entire LISA frequency domain. 

Subsequently, in \cite{heavy_stochastic}, we enhanced the statistical framework discussed above by incorporating an analysis using B-splines following a shape-agnostic spectral model to estimate the PSD of the noise, similarly to~\cite{Baghi:2023qnq}. Furthermore, we characterized stochastic GW signals~\cite{heavy_stochastic} by adopting this framework based on a segmented set of hyperbolic parameters. Specifically, our approach leverages the hyperbolic likelihood to investigate the spectral characteristics of the signal and concurrently check for any deviations from Gaussianity. The analysis focuses on a stochastic signal generated by overlapping signatures in LISA band, expected to be cyclo-stationary due to orbital motion. The study recovers non-stationarity as a deviation from Gaussianity, particularly in a specific frequency range~\cite{heavy_stochastic}. 

The present study aims to go one step further and extend this robust statistical framework to encompass the full frequency domain and simultaneously characterize the noise properties while performing PE for the given GW source. By broadening the applicability of the heavy-tailed likelihood approach, we seek to enhance the overall robustness and reliability of GW PE across a wider range of scenarios and data characteristics. Among the different types of GW sources, we apply this framework for Binary Black-Hole (BBH) mergers, which is a common source for both space-based \cite{Amaro2017,chinesedetectors} and ground-based detectors \cite{LIGO, Virgo,KAGRA,ET,CE,NEMO}, allowing us to test this framework for various GW detectors. Specifically, BBH mergers span a wide range of masses and can be broadly categorized into two groups:
\begin{itemize}
    \item Supermassive BBH mergers: These involve much larger black holes, with masses ranging from millions to billions of solar masses. They are believed to be the result of galaxy mergers and are a primary target for future space-based detectors \cite{Amaro2017,chinesedetectors}.
    
    \item Stellar-mass BBH mergers: These involve black holes with masses typically ranging from a few to tens of solar masses. They are thought to be the end products of massive star evolution and are primarily detected by ground-based interferometers \cite{LIGO, Virgo,KAGRA,ET,CE,NEMO}.
\end{itemize}

Detecting and analyzing BBHs allows scientists to probe fundamental questions in physics and astrophysics (see \cite{mapelli2018} and references therein). 
The first direct detection of GWs from a stella-mass BBH merger by Advanced LIGO in 2015 (GW150914)~\cite{firstGW} opened up an entirely new way of observing the universe. Since then, numerous BBH mergers have been detected \cite{GWTC,GWTC1,GWTC-2,GWTC-4,GWTC-3,GWTC-5, Nitz_2023, Abbott_2023, koloniari2024, GWTC-4}, each contributing to our understanding of these extreme cosmic events (for example, see \cite{Abbott_2021_H,PhysRevX.6.041015,PhysRevD.102.043015,Abbott_2019_3800,Abbott_2016_rate}) and further detections will give insight into the nature of our universe \cite{kalogera2019deeper,Farr_2019}. 

The expected detection rate of GW from BBH is projected to increase significantly in the future as more advanced detectors come online and existing detectors improve their sensitivity. With third-generation detectors~\cite{ET,CE,NEMO}, detection rates could increase to thousands or even millions of events per year, making it very possible to have overlapping low-SNR signals in time and frequency. Similar considerations apply to BNS and BHNS systems, for which the typically longer signal duration in the detector band further increases the probability of signal overlap. For this reason, performing PE of a BBH in a very complex data set could be challenging. Gaussian-like likelihoods are expected to provide biased results, since low-SNR signals would break the Gaussianity of the noise and increase noise uncertainties. We simulate such cases and give more details in Sect.~\ref{sec:LIGO scenario methodology}.

This paper presents two key case studies that demonstrate the advantages of our framework
\begin{enumerate}
    \item Analysis of a Massive Black Hole Binary merger (MBHB) in simulated LISA data, characterized by Gaussian noise (see Sect.~\ref{sec:LISA_methodology}). This case allows us to benchmark the performance of our extended hyperbolic likelihood method against the traditional Whittle likelihood under ideal noise conditions. In addition, in Appendix A, we present a comparison between the Whittle and Gaussian likelihoods.

    \item Examination of a stellar-mass black hole binary merger using real ground-based gravitational-wave detector data, containing non-Gaussian noise features (see Sect.~\ref{sec:LIGO scenario methodology}). This scenario enables us to assess the robustness of our method in handling real-world data complexities and to demonstrate its advantages over conventional approaches.
\end{enumerate}

Through these analyses, we aim to show that this framework maintains high performance under ideal conditions while offering robustness in the presence of non-Gaussian noise and outliers. This advancement has significant implications for improving the accuracy and reliability of gravitational-wave data analysis across various detectors and astrophysical sources.

The paper is structured as follows. Sect.~\ref{sec:framework} provides a detailed description of the statistical framework, including the extension of the hyperbolic likelihood method to the full frequency domain. Sect. \ref{sec:methodology} outlines our methodology, detailing the data sources, simulation setups, and main results. Sect. \ref{sec:conclusions} discusses the implications of our findings for GW astronomy and an outlook on future applications. Appendix~A presents a
benchmark comparison between Gaussian and Whittle likelihoods in the LISA
scenario under controlled noise assumptions, while Appendix~B provides
supplementary details and results for the
multiple-injection and glitch analyses.

\section{Statistical Framework\label{sec:framework}}
In a Bayesian framework, the posterior distribution of the parameters $\param$ is given by
\begin{equation}
p(\param | x) \propto p(x|\param)p(\param),
\label{eq:bayes}
\end{equation}
where $ p(x|\param)$ represents the likelihood function and $p(\param)$ denotes the prior distribution of the parameters. The marginal likelihood $p(x)$ typically serves as a normalization constant and is therefore omitted from Eq.~(\ref{eq:bayes}), although it plays a central role in model selection~\cite{gelmanbda04}.

\subsection{Data model and residuals in GW analysis}

In GW data analysis, detector data are modeled as the sum of signal and noise,
\begin{equation}
s(t) = h(t;\param_{\mrm{source}}) + n(t;\paramS),
\end{equation}
where $h$ is the GW signal depending on source parameters $\param_{\mrm{source}}$, and $n$ denotes instrumental noise, which may be described by a set of noise parameters $\paramS$. 

Given a candidate signal model, the residual data are defined as
\begin{equation}
x(t) = s(t) - h(t;\param_{\mrm{source}}),
\end{equation}
which represent the noise realization under the assumed signal hypothesis. Likelihoods are constructed from these residuals.

In the GW community, data analysis is commonly performed in the frequency domain, where stationary colored noise is naturally described by the PSD. In this representation, the PSD defines the metric used to measure residual energy through the noise-weighted inner product.

\subsection{Noise-weighted inner product and Gaussian likelihood}

The noise-weighted inner product between two real time series $a$ and $b$ is defined as~\cite{PhysRevD.69.082005, PhysRevD.49.2658}
\begin{equation}
\langle a | b \rangle = 4 \, \mathrm{Re}\int\limits_0^\infty \mrm{d}f \left[ \tilde{a}^\ast(f) C_n^{-1}(f) \tilde{b}(f) \right],
\label{eq:inprod}
\end{equation}
where $C_n(f)$ is the one-sided noise PSD, the tilde $(\,\tilde{ }\,)$ denotes the Fourier transform, and $(\,^\ast\,)$ denotes complex conjugation. For multiple data channels or detectors, $C_n$ becomes a matrix whose off-diagonal components describe cross-spectral noise correlations~\cite{Sasli:2023mxr, heavy_stochastic}.

The quadratic residual energy appearing in likelihood constructions can then be written as
\begin{equation}
r_i = {\rm Re}\, \left\{ \tilde{x}_i^\ast C_n^{-1} \tilde{x}_i \right\},
\label{eq:residualsf}
\end{equation}
which represents the noise-weighted residual power in the frequency domain.

In this framework, the Gaussian log-likelihood is commonly written as
\begin{equation}
\Lambda_\mrm{\cal N}(\param_{\mrm{source}}) \propto -\frac{1}{2}\langle x | x\rangle.
\label{eq:gaussian_llh}
\end{equation}

The Whittle approximation of the Gaussian likelihood (see, e.g., \cite{Cornish:2020dwh, LISACosmologyWorkingGroup:2022jok,Flauger:2020qyi,Contaldi:2020rht,Armano:2018ucz,Sasli:2023mxr}) takes the form
\begin{equation}
\Lambda_{\cal W}(\param) \propto - \sum_f  \left( \ln C (\param)  + \langle x | x \rangle \right),
\label{eq:whittle}
\end{equation}
with $\param \equiv \{\param_{\mathrm{source}}, \paramS\}$, where $C(\param)$ denotes the parametrized noise covariance and $\paramS$ the PSD parameters.

\subsection{Hyperbolic likelihood formulation}
To address limitations associated with likelihood models that assume Gaussian noise properties, one may adopt likelihood models with heavier tails. Several approaches have been proposed in the literature, including likelihoods based on Student's $t$ distribution~\cite{roever2011A, roever2011B}, or mixture-based likelihood constructions~\cite{PhysRevD.60.021101, PhysRevD.65.122002}. Such approaches aim to provide robustness against non-Gaussian noise features and transient deviations from stationarity.

In this work, we adopt the Hyperbolic likelihood model~\cite{Prause1999TheGH, Eberlein2002TheGH, Sasli:2023mxr, heavy_stochastic}. One of the key advantages of this model is its flexibility in describing a wide range of statistical behaviors in the data, including deviations from Gaussianity, while retaining a well-defined likelihood structure. As demonstrated in~\cite{Sasli:2023mxr, heavy_stochastic}, the \hyp{} likelihood is sensitive to departures from Gaussian noise assumptions and can capture heavy-tailed residual statistics. By jointly estimating the parameters of the \hyp{} likelihood together with the physical model parameters, we gain additional information about the statistical properties of the residual data (see Eq.~(\ref{eq:residualsf})).

For data vectors $\bm{x}_i \in \mathbb{R}^d$, $1 \le i \le n$, where $d$ denotes the dimensionality of the data representation adopted in the likelihood (i.e., the number of real degrees of freedom per sample), the symmetric Hyperbolic log-likelihood is written as~\cite{Sasli:2023mxr}
\begin{equation}
\begin{aligned} 
\Lambda_{\cal H}(\alpha, \delta) = & n\Bigg[ \frac{d+1}{2} \ln\left(\frac{\delta}{\alpha}\right) + \frac{1-d}{2} \ln(2\pi)  \\
    & - \ln(2\alpha)  - \ln \left[K_{(d+1)/2} (\delta\alpha)\right]\Bigg] \\
    & - \alpha\sum^n_{i=1}\sqrt{\delta^2 + r_i},
\end{aligned}
\label{eq:hyp}
\end{equation}

where
\begin{equation}
r_i\equiv\bm{x}_i^{\rm T} \widehat{\Delta}^{-1} \bm{x}_i,
\end{equation}
and $\widehat{\Delta}$ denotes an estimate of the symmetric positive definite dispersion matrix defining the quadratic metric of the distribution. In the Gaussian limit, this matrix reduces to the noise covariance, or equivalently the PSD operator in the frequency domain.

The analytic expression for the variance of the symmetric Hyperbolic distribution is given by~\cite{Sasli:2023mxr}
\begin{equation}
\sigma^2 = \frac{\delta K_{2}(\delta\abs{\alpha})}{\alpha K_{1}(\delta\abs{\alpha})},
\label{eq:variance}
\end{equation}
where $K_1$ and $K_2$ are modified Bessel functions of the third kind. In the Gaussian limit, $\delta,\alpha \rightarrow \infty$ such that $\delta/\alpha \rightarrow \sigma^2$. In the univariate time-domain formulation, $\sigma^2$ reduces to the noise variance, whereas in the frequency-domain GW formulation this generalizes to the frequency-dependent noise PSD \cite{Sasli:2023mxr, Sasli:2025tjf}.

In gravitational-wave applications, precisely defining the dimensionality $d$ is important. For $n_c$ independent data channels (detectors), we take $d=n_c$ in the time domain. In the frequency domain, each channel contributes one complex Fourier coefficient, corresponding to two real degrees of freedom, and therefore we set $d=2n_c$ per frequency bin~\cite{heavy_stochastic}.

Depending on the frequency range considered in the analysis, we may adopt a piecewise parametrization of the Hyperbolic model in order to achieve a robust reconstruction of the effective noise PSD. Specifically, the frequency domain can be divided into $N_s$ frequency bins, within which the Hyperbolic parameters are inferred independently. Therefore, the parameters can be taken to be ${\alpha_{N_s}, \delta_{N_s}}$, ${\alpha, \delta_{N_s}}$, or ${\alpha_{N_s}, \delta}$. For computational efficiency, in this work we adopt the ${\alpha, \delta_{N_s}}$ parametrization.

In conventional GW parameter estimation, one may introduce explicit noise model parameters $\paramS$ to reconstruct the PSD of the noise, leading to an inferred parameter set $\param\equiv \{\param_{\mrm{source}}, \paramS\}$. In contrast, when adopting the \hyp{} likelihood, only minimal assumptions about the detailed noise model are required, and inference is instead performed directly on the parameters of the \hyp{} distribution. In this way, the Hyperbolic parameters provide an effective statistical description of the residual noise properties while simultaneously enabling robust parameter estimation of the physical source model.

\section{Analysis \label{sec:methodology}}
Our previous studies demonstrated the robustness of the hyperbolic likelihood in handling both Gaussian and non-Gaussian noise. For Gaussian noise with a single data source, both the \hyp{}  and \whtl{} likelihoods performed similarly regardless of the noise modeling accuracy, while the Gaussian likelihood was highly dependent on the correct PSD of the noise \cite{Sasli:2023mxr}. In non-Gaussian noise conditions, the \hyp{}  likelihood accurately recovered the noise variance, in contrast to the \whtl{} likelihood, which overestimated it \cite{heavy_stochastic} due to its Gaussian noise assumption. In addition, the importance of extracting the underlying statistical properties of the data is highlighted in \cite{heavy_stochastic}. For example, the detection and the characterization of non-Gaussian stochastic signals from ultra-compact binaries could hinder the underlying physical processes and the population of these sources in the universe~\cite{heavy_stochastic}.

For the purposes of this paper, we adopt a combined methodology from our previous studies. We perform PE for BBH under both ideal (Sect.~\ref{sec:LISA_methodology}) and non-ideal (Sect.~\ref{sec:LIGO scenario methodology}) noise conditions, following the statistical framework described in Sect.~\ref{sec:framework}, while accounting for corrections in the assumed PSD noise. We assume that the data channels are statistically uncorrelated, such that the total log-likelihood factorizes into the sum of the log-likelihoods of the individual channels. This approximation is well justified for the datasets considered here. For LISA, we analyze the approximately independent A and E TDI channels~\cite{Tinto:2004wu, snrtn}, while for ground-based observations we treat data from independent detectors, such as LIGO Livingston (L1) and Hanford (H1), as uncorrelated.

The PE with hyperbolic likelihood \hyp{}  is compared with Whittle \whtl{} and Gaussian likelihood. For the \whtl{} likelihood, we assume a simple noise model; a level correction on the assumed PSD of the noise, $\paramS$. An alternative case could be the assumption of the actual LISA noise model, and infer the nominal values for the SciRD noise, for example see Appendix~\ref{sec:appendix A}. The dimensionality of this parameter depends on the number of frequency bins $N_s$ we divide into the frequency domain. Therefore, the noise parameters correspond to band-dependent PSD correction factors. For the \hyp{} likelihood, we introduce band-dependent $\delta_{N_s}$ parameters, while assuming a common $\alpha$ parameter across the full frequency band, consistent with the formulation described in Sect.~\ref{sec:framework}. This results in $N_s$ additional noise parameters for the \whtl{} likelihood and $N_s+1$ for the \hyp{} likelihood.

\subsection{Sampling strategy}

For PE purposes, we employ two packages, \texttt{Eryn}~\citep{eryn} and \texttt{pocoMC} \cite{karamanis2022pocomcpythonpackageaccelerated}.
\texttt{Eryn} utilizes a Parallel Tempering Markov Chain Monte Carlo (PTMCMC) approach~\citep{eryn, emcee}, which is particularly effective for gravitational wave inference problems where posterior distributions may exhibit strong correlations, non-Gaussian structures, and multi-modality. By simultaneously evolving multiple ensembles of Markov chains at different inverse temperatures, \texttt{Eryn} facilitates efficient exploration of complex likelihood landscapes. The exchange of chains at neighboring temperatures helps low-temperature chains escape local maxima, thereby improving convergence and robustness. Initialized with walkers drawn from small perturbations around the injected parameters, \texttt{Eryn} accelerates burn-in while ensuring thorough exploration of the posterior. Proposals are generated using affine-invariant ensemble moves, with samples from the lowest-temperature chain representing the target posterior distribution, providing a reliable baseline for posterior sampling.

\texttt{pocoMC} is a Preconditioned Monte Carlo sampler~\citep{karamanis2022pocomcpythonpackageaccelerated} that combines annealed Persistent Sampling~\citep{karamanis2025persistent} with normalizing flows, which are trained on-the-fly to learn an approximate transport map from the target distribution to a latent space with reduced correlations. This method significantly enhances sampling efficiency compared to traditional MCMC or nested sampling methods. The algorithm evolves a population of particles through a sequence of intermediate distributions, updating the normalizing flow to effectively ``Gaussianize" the distribution and enable large, informed proposal steps. \texttt{pocoMC} has been validated on gravitational wave inference problems using \texttt{bilby}, demonstrating substantial speedups and accurate posterior reconstruction~\citep{Williams_2025, Vretinaris_2026}. In this work, we further improve \texttt{pocoMC}’s performance with a fully vectorized likelihood implementation, making it particularly well-suited for computationally demanding analyses with complex likelihoods.

\subsection{LISA scenario: Simulated Gaussian noise\label{sec:LISA_methodology}}

\begin{ruledtabular}
\begin{table}[t]
\centering
    \begin{tabular}{lc}
       \textbf{Parameters} & SNR = 357 \\
      \hline
    Mass 1, $m_1~[M_\odot]$ & 4956676.2876  \\ 
    Mass 2, $m_2~[M_\odot]$ & 4067166.60352   \\
    Spin of body 1, $\chi_1$ & -0.523732715 \\
    Spin of body 2, $\chi_2$ & -0.117412144 \\
    Distance, D [Mpc] & 61097.116076  \\ 
    Inclination, $\iota~[rad]$  & 1.420048341  \\ 
    Ecliptic Latitude, $\beta~[rad]$ & -1.081082148 \\
    Ecliptic Longitude, $\lambda~[rad]$ & 4.052962883  \\
    Polarization, $\psi$ & 1.22844350008 \\
    Phase at coalescence, $\phi_0$ & 0.6417162631 \\
    \end{tabular}
    \caption{Parameters of our model from Sangria training data-set. The SNR of this case is 357 for a one-year data. \label{tab:models}}
\end{table}
\end{ruledtabular}

We investigate a low-SNR case from the LISA Data Challenge (LDC) {\it Sangria} data set~\cite{ldc}, corresponding to a massive black hole binary (MBHB) signal. Waveforms are generated using the {\tt \textbf{BBHx}} package~\citep{Katz_2020,Katz_2022,michael_katz_2021_zenodo} with the {\tt PhenomD} approximant~\cite{Khan:2015gcj}. The injected waveform parameters are listed in Table~\ref{tab:models}. The simulated data have a total observation duration of $T_\mathrm{obs}=30.4368~\mathrm{days}$, while the coalescence time is set to one day before the end of the measurement.

The inferred source parameters $\param_{\mrm{source}}$ correspond to those listed in Table~\ref{tab:models}. However, instead of sampling directly in the component masses $(m_1, m_2)$, we sample in the chirp mass $\mathcal{M}$ and mass ratio $\mrm{q}$ $(\equiv m_1/m_2)$. 

\par

\begin{table}[t]
\centering
\caption{The prior settings for the parameters $\param_{\mrm{source}}$.}
\label{tab:priors}
\begin{tabular}{c@{\hskip 0.5in}c@{\hskip 0.5in}c}
\hline
\hline
$\bm{\param_{\mrm{source}}}$ & \textbf{Lower bound} & \textbf{Upper bound} \\
\hline
$\mathcal{M}$ & $0.39\cdot 10^{6}$ & $7.8\cdot 10^{6}$ \\
$\mrm{q}$ & 0.99999 & 2 \\
$\chi_1$ & -1 & 1 \\
$\chi_2$ & -1 & 1 \\
$\log_{10}\mrm{D}$ & $0.1\log_{10}\mrm{D_{true}}$ & $2\log_{10}\mrm{D_{true}}$ \\
$\mrm{t_c}$ & 365.2422 & 8765.8128 \\
$\cos\iota$ & -1 & 1 \\
$\sin\beta$ & -1 & 1 \\
$\lambda$ & 0 & $2\pi$ \\
$\psi$ & 0 & $\pi$ \\
$\phi_0$ & 0 & $2\pi$ \\
\hline
\hline
\end{tabular}
\end{table}

\begin{figure*}[t]
  \centering
  \begin{subfigure}[t]{0.7\textwidth}
    \centering
    \includegraphics[width=\linewidth]{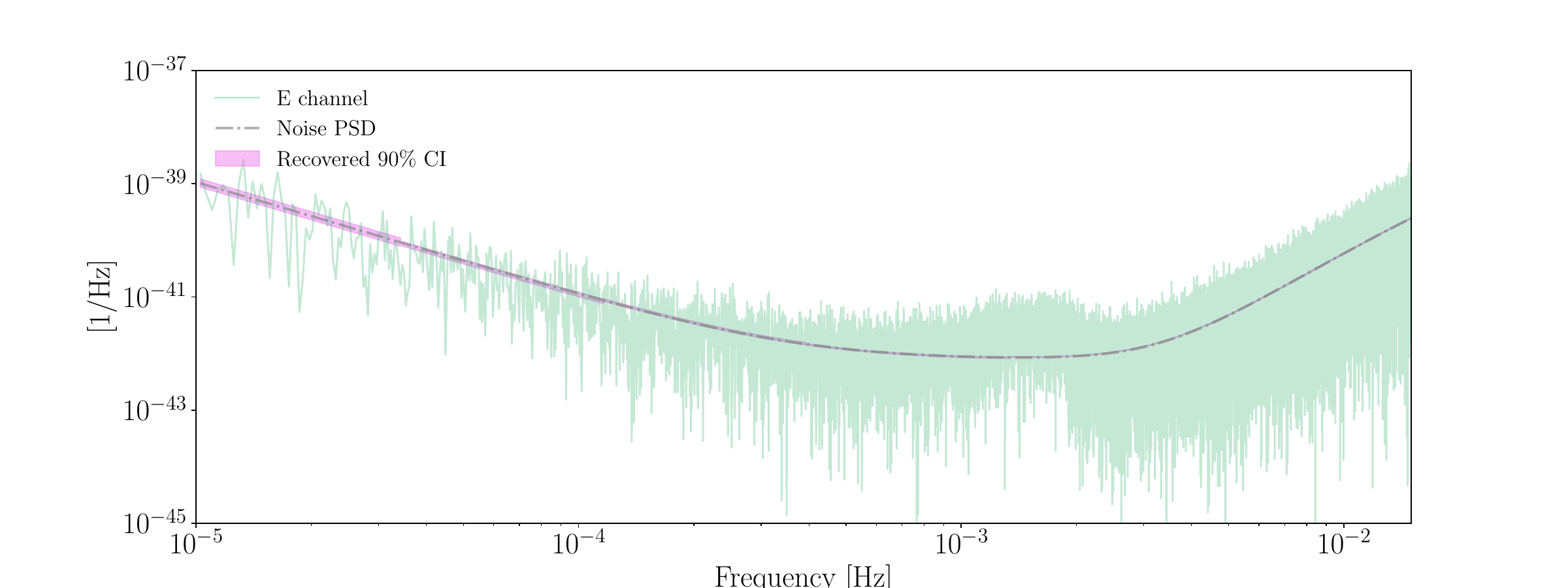}
    \caption{Recovered PSD in the LISA band using the \hyp{}  likelihood.}
    \label{fig:LISA_recovered_psd}
  \end{subfigure}\hfill
  \begin{subfigure}[t]{0.3\textwidth}
    \centering
    \includegraphics[width=0.85\linewidth]{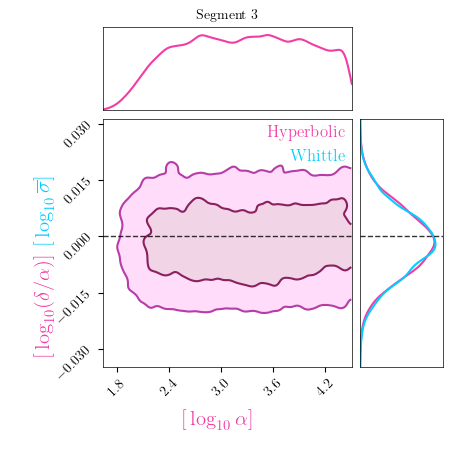}
    \caption{Example of PSD correction for a randomly selected frequency segment. Segment 3 corresponds to the third frequency bin used in the analysis and is shown with the \hyp{} (pink) and \whtl{} (blue) likelihoods. Both predict no correction to the adopted PSD.}
    \label{fig:LISA_seg3}
  \end{subfigure}
  \caption{PSD analysis in the LISA band.}
  \label{fig:LISA_psd_combined}
\end{figure*}

Considering the LISA constellation and instrument, we assume a rigid constellation, and equal noise across all space-crafts. This allows us to use the noise-orthogonal set of TDI channels ($A,\,E$ and $T$)~\cite{TDI, lisa_optimal_sen}, computed from the $X, \,Y, \, Z$ channels as
\begin{equation}
\begin{aligned}
    A = \frac{1}{\sqrt{2}}(Z - & X), \quad E = \frac{1}{\sqrt{6}}(X - 2Y + Z), \\
    T &= \frac{1}{\sqrt{3}}(X + Y + Z).
    \label{eq:aet}
\end{aligned}
\end{equation}
The analytic expression for the average noise level in the $A$, $E$ TDI channel is
\begin{eqnarray}
PSD^{\mrm{A}\mrm{E}}(f) = & 8\, \mrm{sin}(\hat{\omega})^2 \big\{ 2 S_\mrm{i} \left[ 3 + 2\mrm{cos}(\hat{\omega})^2 + \mrm{cos}(2\mathcal{\hat{\omega}}) \right] \nonumber\\
& + S_\mrm{a} \left[ 2+\mrm{cos}(\hat{\omega})\right]  \big\},
\label{eq:s_psd_model}
\end{eqnarray}
where $\hat{\omega}= 2\pi f L/c$, $f$ the frequency and L is the average detector arm-length. The noise components $S_i$ and $S_a$ are fixed for the general case and selected to be in agreement with the nominal values for the SciRD noise~\citep{SciRD,Flauger_2021},
\begin{equation}
\begin{aligned}
    S_a &= \mrm{log}_{10}\left(3\cdot 10^{-15}\right)^2 \quad \frac{\mrm{m}^2}{\mrm{Hz}\cdot \mrm{sec}^4}, \\
    S_i &= \mrm{log}_{10}\left(15\cdot 10^{-12}\right)^2 \quad \frac{\mrm{m}^2}{\mrm{Hz}},
    \label{eq:sa_si_values}
\end{aligned}
\end{equation}
which are expressed in relative frequency units~\cite{snrtn}. For this simple investigation, the $T$ channel is neglected since it greatly suppresses the GW signal.

We run the PTMCMC with 20 temperatures, 80 walkers, and $3\times10^5$ samples per walker, while we set the burn-in samples equal to $1\times10^5$. The inferred parameters are the source parameters and the noise fitting parameters for $N_s$=6 equally in log spaced frequency bins. This results in a total parameter space $\param$ of 19 when account for \whtl{} likelihood, and 20 for \hyp{}  likelihood.

Finally, we have adopted wide priors for the source parameters, which are summarized in Table~\ref{tab:priors}. The parameters with which we perform the noise fitting as well as the first eight source parameters follow the uniform distribution, and the remaining follow a periodic prior. The whole PE is performed on GPUs, using cuda~\cite{cuda}, increasing the efficiency up to 10 times compared to CPU-based analyses.

\begin{figure*}[t]
 \includegraphics[width=0.9\linewidth]{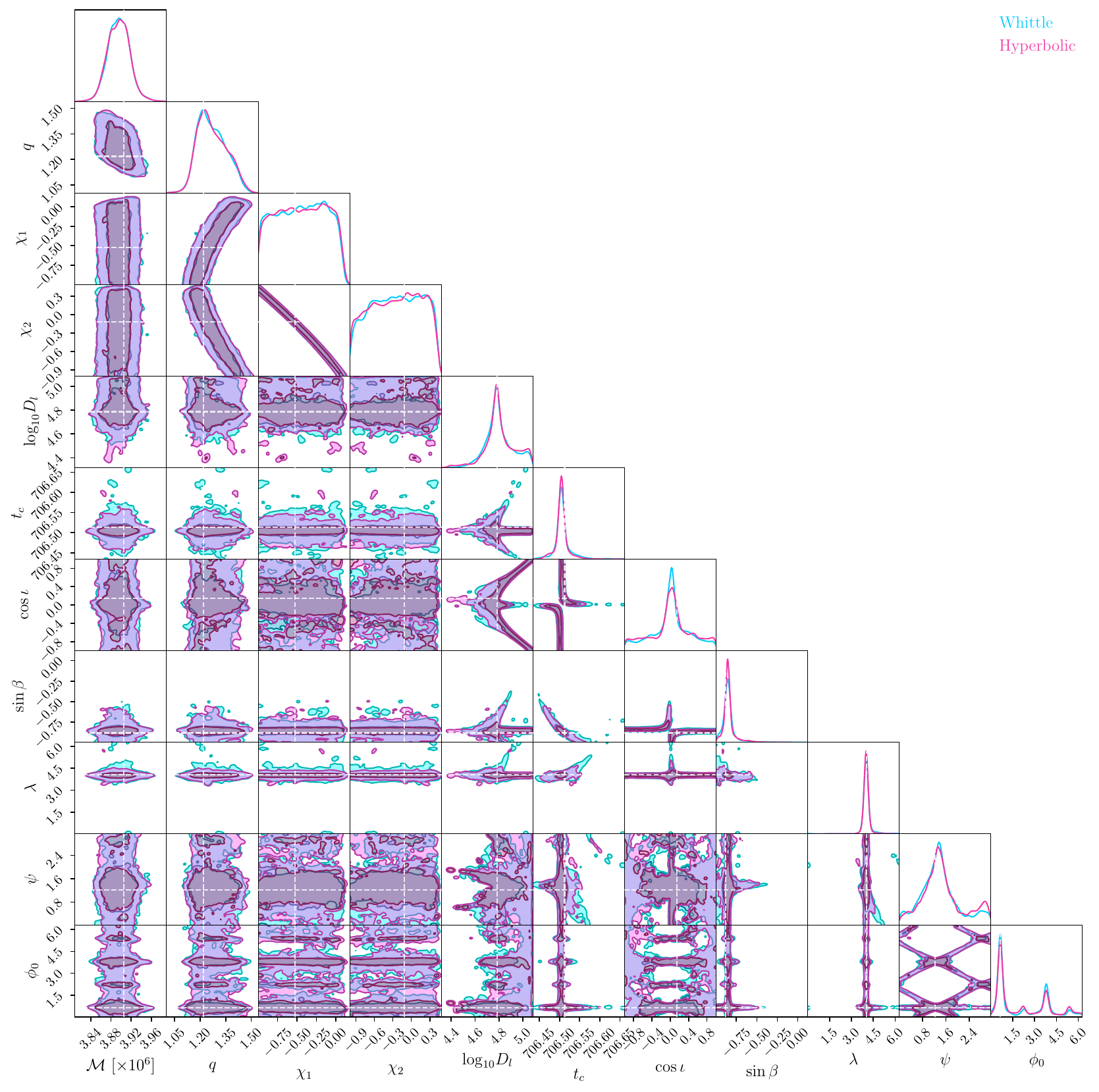}
 	\caption{Cornerplots using the \whtl{} (blue) and the \hyp{} (pink) likelihood. This case correspond to a source of 357 SNR for one-year data (see Table~\ref{tab:models}). Units follow Table \ref{tab:models}.}
 \label{fig:LISA_posterios}
\end{figure*}

Figure~\ref{fig:LISA_recovered_psd} illustrates the PSD reconstruction for the LISA scenario. The plot shows the true noise PSD (black line) along with the recovered $90\%$ confidence interval (shaded purple region) obtained using the \hyp{} likelihood. Since both likelihoods converge to the Gaussian limit in this regime, the reconstructed PSDs from the \hyp{} and \whtl{} likelihoods are statistically indistinguishable, and we therefore show only the \hyp{} reconstruction for clarity. Both likelihood methods demonstrate excellent performance in reconstructing the PSD across the frequency range of interest $(10^{-5}$ to $10^{-2}$ Hz). The recovered PSD closely follows the true noise PSD, with the $90\%$ confidence interval encompassing the true PSD line over most of the frequency range.

In Fig.~\ref{fig:LISA_seg3}, we show the PSD correction posteriors for a representative frequency segment. The posteriors obtained using both likelihoods indicate that no significant refinement of the PSD is required in this segment. The distributions for both methods (pink for \hyp{} and blue for \whtl{}) are relatively compact and centered near the true value, corresponding to zero logarithmic PSD correction.

We further compared the performance of the \hyp{}  likelihood method with the traditional \whtl{} likelihood. The results demonstrate that both methods perform comparably well in this scenario, characterized by Gaussian noise.
Fig. \ref{fig:LISA_posterios} presents the posterior distributions for various parameters of the binary system. The overlapping posterior distributions for the \hyp{}  and \whtl{} likelihoods indicate that both methods provide consistent PEs. This similarity in performance suggests that the proposed \hyp-likelihood analysis converges to a likelihood shape that closely resembles the Gaussian, thus the similar results to the whtl{} likelihood analysis.

\subsection{Ground-based detector scenario: Real noise\label{sec:LIGO scenario methodology}}

\begin{ruledtabular}
\begin{table}[t]
\centering
    \begin{tabular}{lc}
    \multirow{2}{*}{\textbf{Parameters}} & {$\bm{\mrm{SNR_{L1}}}=24$}\\
        & $\bm{\mrm{SNR_{H1}}}=14$ \\
      \hline
    Mass 1, $m_1~[M_\odot]$ & 36  \\ 
    Mass 2, $m_2~[M_\odot]$ & 29   \\
    Distance, D [Mpc] & 1000  \\
    Right Ascension (RA) & 1.375 rad \\
    Declination (Dec) & 0.2108 rad  \\
    Polarization, $\psi$ & 2.659 rad \\
    Phase at coalescence, $\phi_0$ & 1.3 rad \\
    Inclination angle, [rad]  & 0.4  \\
    Primary object tilt [rad] & 0 \\
    Secondary object tilt [rad]  & 0 \\
    Primary dimensionless spin magnitude & 0\\
    Secondary dimensionless spin magnitude & 0\\
    Relative spin azimuthal angle, $\phi_{12}$ [rad] & 0\\
    Spin phase angle, $\phi_{jl}$ [rad] & 0\\
    \end{tabular}
    \caption{Parameters of our targeted BBH injection in LIGO data. \label{tab:model ligo}}
\end{table}
\end{ruledtabular}

We investigate three different cases of real noise, but with more complex scenarios. These cases consist of datasets where we have a targeted BBH in the presence of (1) seven sub-threshold SNR BBHs, (2) a long-lasting low-SNR BBH, and (3) a noise transient (glitch). For each dataset, the PSD is estimated from 128 s of data using a Welch-style approach with 8 s Fourier segments windowed with a Tukey window ($\alpha=0.2$) and 50\% overlap, yielding 31 segments. Individual periodograms are computed for each segment and combined using the median at each frequency bin to obtain a robust PSD estimate.

For BBH waveform modeling and signal injections into real data, we employed the \texttt{bilby} package \cite{bilby_paper} and adopted \texttt{IMRPhenomPv2}~\cite{Pv2_wp}. This allows us to simulate gravitational-wave signals within the context of actual detector noise, providing a realistic scenario for testing our analysis methods. For every case, we inject the same signal, a simulated BBH with parameters as shown in Table \ref{tab:model ligo}. Those parameters were chosen to represent a typical stellar-mass BBH system detectable by ground-based detectors.

When running with \texttt{Eryn}, we use 20 temperatures, 50 walkers, and $7\times10^4$ samples per walker, with $5\times10^4$ burn-in samples. For \texttt{pocoMC}, we set $n_{\rm eff}=2000$, $n_{\rm active}=1000$, and use $5\times10^4$ total samples, corresponding to a conservative configuration. In this setup, \texttt{pocoMC} achieves approximately a fourfold speed-up compared to \texttt{Eryn}.

All parameter estimation runs are performed on CPUs, with waveform and likelihood computations parallelized across approximately 70\% of the available cores (32 CPUs), resulting in an effective speed-up of up to a factor of three.

\subsubsection{Multiple sources of BBH \label{sec:multiple sources of BBH}}

We selected a segment of Advanced LIGO data centered on GPS time
\texttt{1238189526.951953}, which is close to a glitch and thus the data contain non-stationary noise when the PSD is calculated. We then investigate two different cases, that could make our PE challenging. The cases are as follows:

\begin{itemize}

    \item \textit{Injecting a long lasting BBH source:}
     We generate a sub-threshold BBH signal (see parameters in Table~\ref{tab:model_long_ligo}) with a duration of 5.7 seconds, which exceeds the 4-second data segment used in our analysis. The long-duration BBH is injected 0.09 seconds prior to the trigger time of the “target” BBH event. This configuration poses a challenging scenario for likelihood formulations that do not account for PSD uncertainty, since the extended signal contaminates the data segment and effectively alters the stationary noise and background properties.

     \item \textit{Injecting seven different low SNR BBH sources}:
    We performed injections of seven low-SNR BBH signals with a common trigger time of \texttt{1238189542.951953}, and the waveform parameters were randomly generated (and Table~\ref{tab:multiple injcetions}). The choice of a common trigger time is intentional. By aligning all sub-threshold BBH signals with the trigger time of the targeted BBH event, we construct a controlled scenario in which multiple weak signals coherently overlap within the same data segment. This setup effectively produces an artificial increase in the apparent signal content at the trigger time, creating a challenging environment for likelihood models. In particular, it allows us to probe the robustness of different likelihood formulations under conditions where the background noise is contaminated by unresolved overlapping signals.

\end{itemize}

\begin{figure}[t]
  \subfloat[Hyperbolic (blue) and Gaussian (yellow) likelihood]{%
\includegraphics[clip,width=0.49\columnwidth]{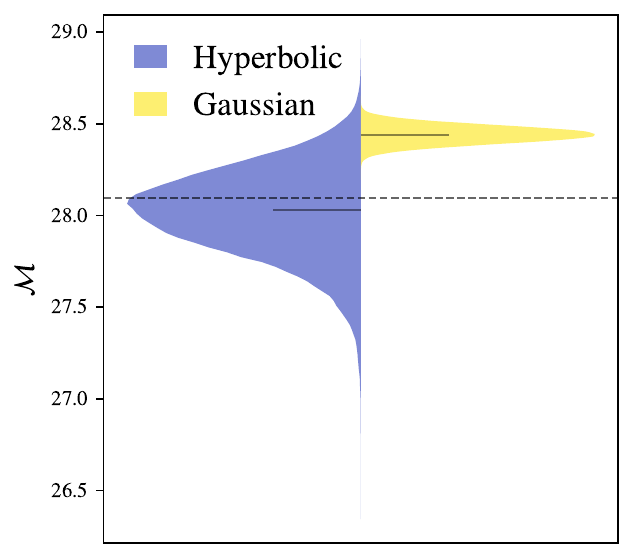}%
\includegraphics[clip,width=0.49\columnwidth]{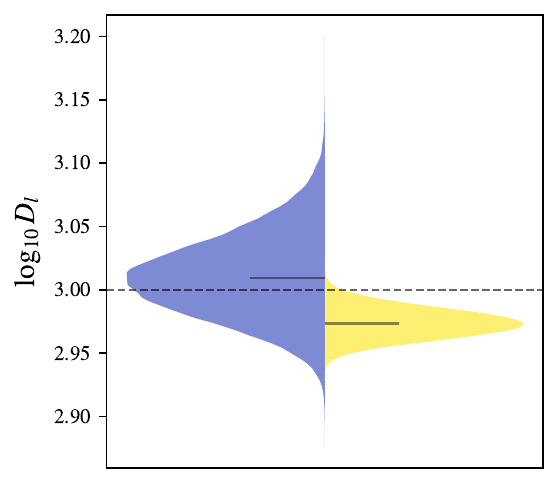}
}

    \subfloat[Hyperbolic (blue) and Whittle (yellow) likelihood]{%
\includegraphics[clip,width=0.49\columnwidth]{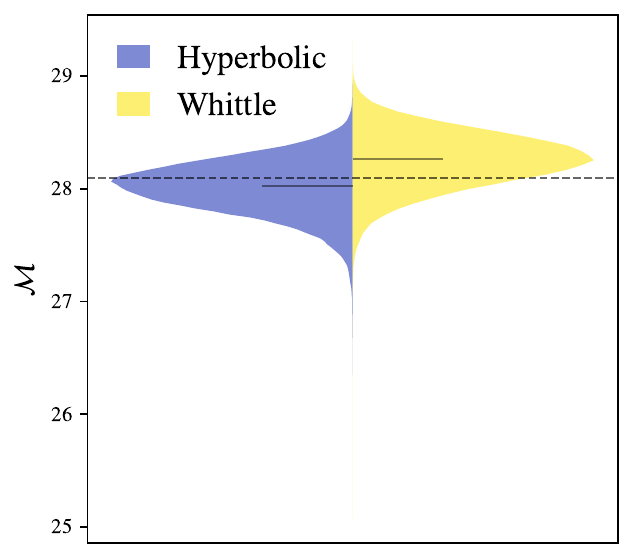}%
\includegraphics[clip,width=0.49\columnwidth]{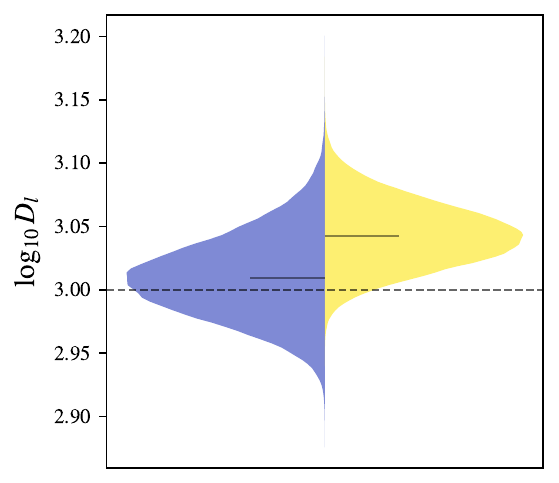}
}

 	\caption{Posterior distributions obtained with the Hyperbolic compared to (a) Gaussian and (b) Whittle likelihoods. Each panel shows split violin plots for the most affected key binary parameters, namely chirp mass $\mathcal{M}~[M_{\odot}]$ and luminosity distance. This case refers to the data with the long-lasting BBH injected. The hyperbolic likelihood systematically reduces bias. 
}
 \label{fig:violin_long}
\end{figure}

\begin{figure}[t]
  \subfloat[Hyperbolic (blue) and Gaussian (yellow) likelihood]{%
\includegraphics[clip,width=0.5\columnwidth]{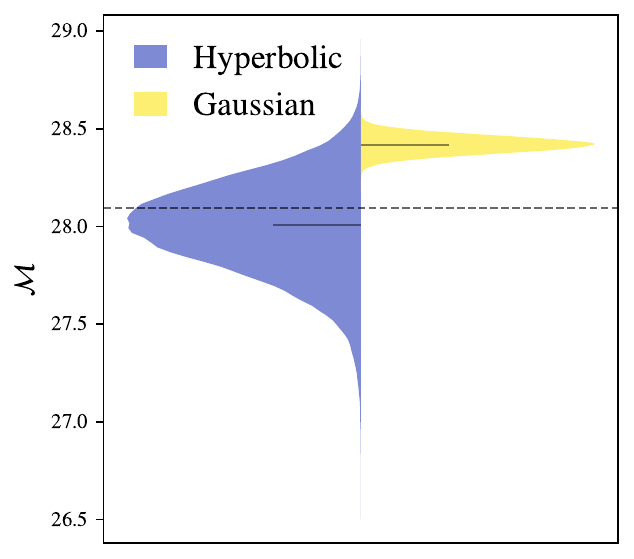}%
\includegraphics[clip,width=0.5\columnwidth]{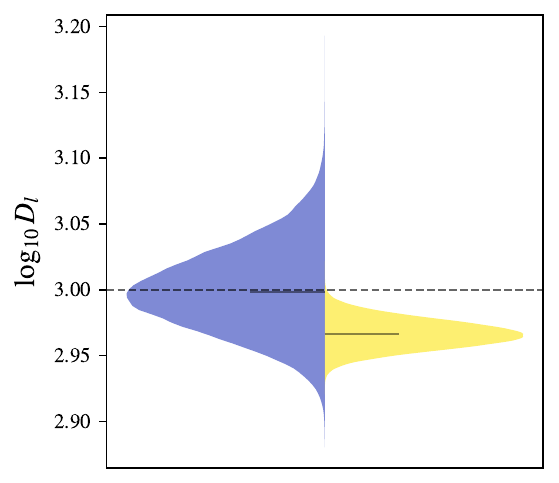}
}

    \subfloat[Hyperbolic (blue) and Whittle (yellow) likelihood]{%
\includegraphics[clip,width=0.49\columnwidth]{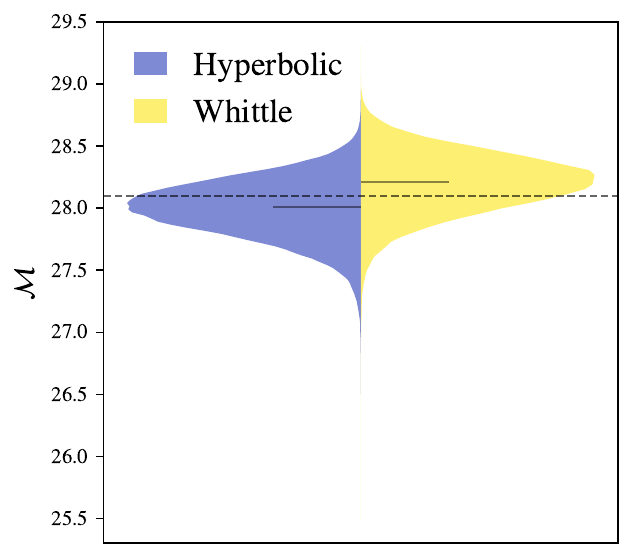}%
\includegraphics[clip,width=0.49\columnwidth]{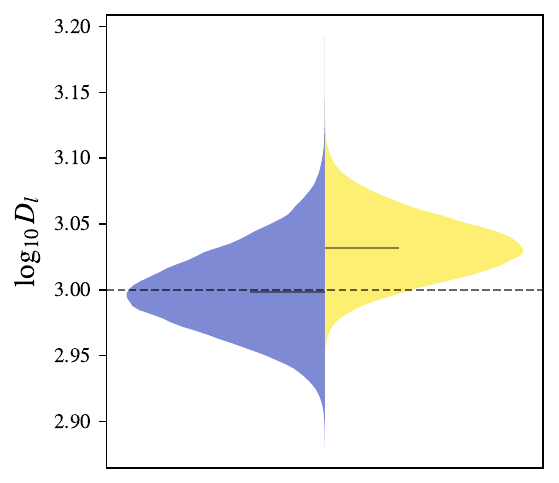}
}

 	\caption{Similar to Fig. \ref{fig:violin_long}, but for the case with the seven low-SNR BBH injections..
}
 \label{fig:violin_multi}
\end{figure}

Figures \ref{fig:violin_long} and~\ref{fig:violin_multi} demonstrate that the \hyp{}  likelihood reduces the bias for the key binary parameters, chirp mass $\mathcal{M}~[M_{\odot}]$ and luminosity distance. Posterior distributions for these cases and waveform reconstructions in the frequency domain are presented in Appendix B with Figures \ref{fig:LIGO_posterios_long_and_multi_bbh}, \ref{fig:FD_recon_long} and \ref{fig:FD_recon_multi}.

\subsubsection{Glitches}
\begin{figure}[h]
    \centering
    \includegraphics[width=1\columnwidth]{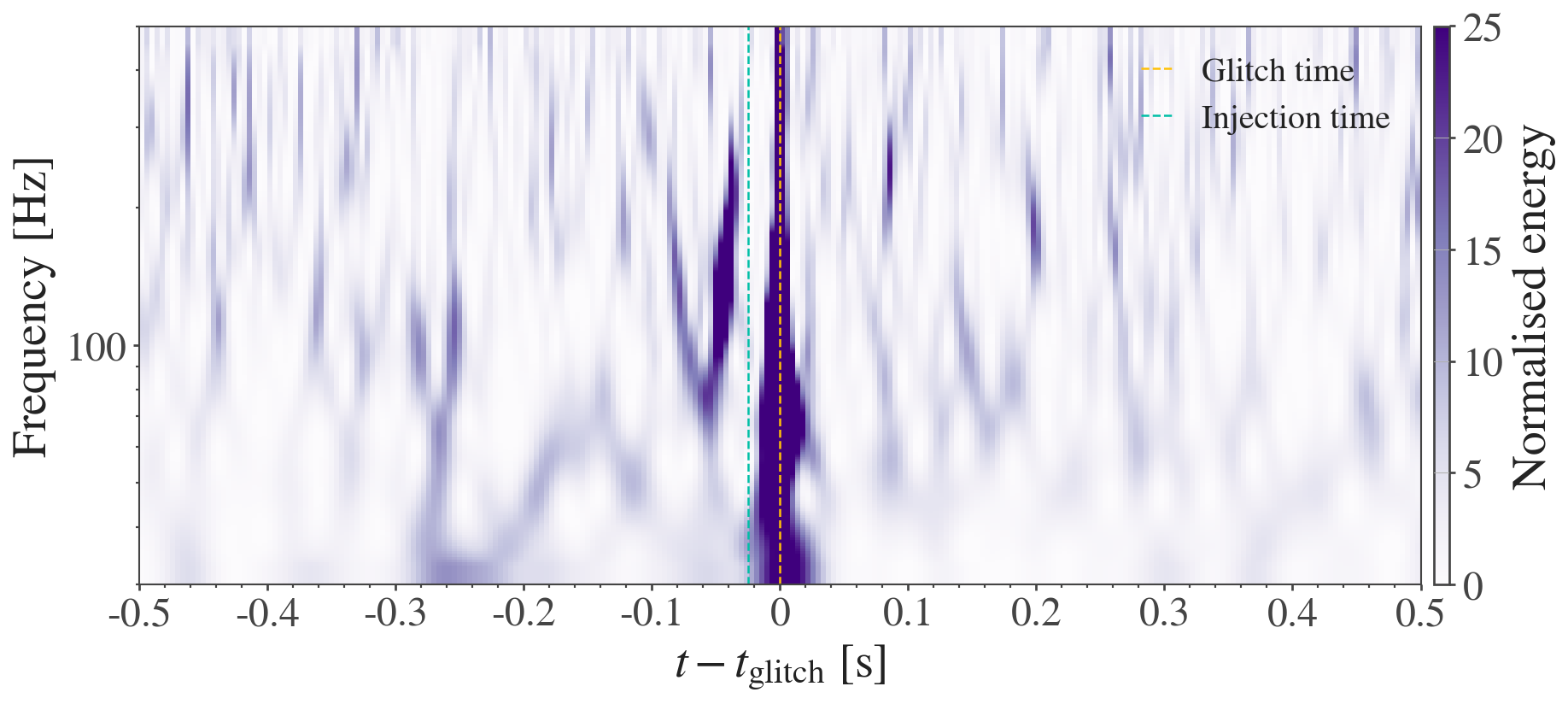}
    \caption{Q-plot for \texttt{1165578732.45} trigger time with the BBH injection. The red dashed line correspond to the glitch time and the green one to the injected signal.}
    \label{fig:qplot}
\end{figure}

In Ref.~\cite{mitigating_glitches}, both successful and unsuccessful cases of PE in the presence of glitches were demonstrated using the \texttt{BayesWave} pipeline. Here, we deliberately focus on a representative case in which the PE remains biased despite the application of glitch-mitigation techniques. Specifically, we consider the ``blip'' glitch with trigger time \texttt{1165578732.45}, where a BBH signal is injected 0.025~s prior to the glitch onset. For this configuration, the method presented in Ref.~\cite{mitigating_glitches} fails to recover the injected parameters: the true values are excluded at the 90\% credible level in both the pre-deglitching and post-deglitching posterior distributions. Fig. \ref{fig:qplot} shows the Q-plot for this data segment from H1 channel (LHO detector) with the injected BBH from Tab. \ref{tab:model ligo}.

\begin{figure}[t]
  \subfloat[Hyperbolic (pink) and Gaussian (green) likelihood]{%
\includegraphics[clip,width=0.95\columnwidth]{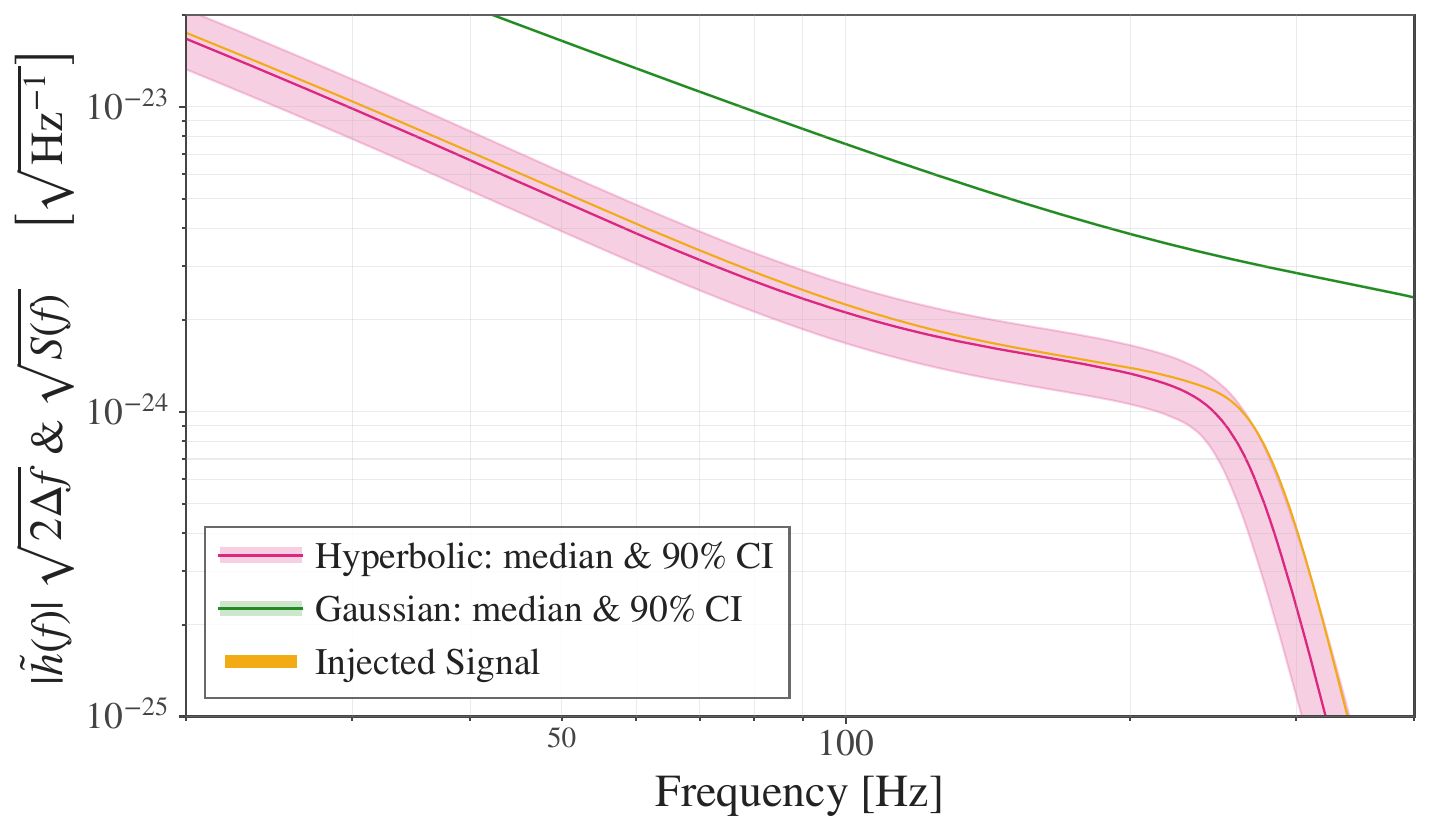}%
}

    \subfloat[Hyperbolic (pink) and Whittle (blue) likelihood]{%
\includegraphics[clip,width=0.95\columnwidth]{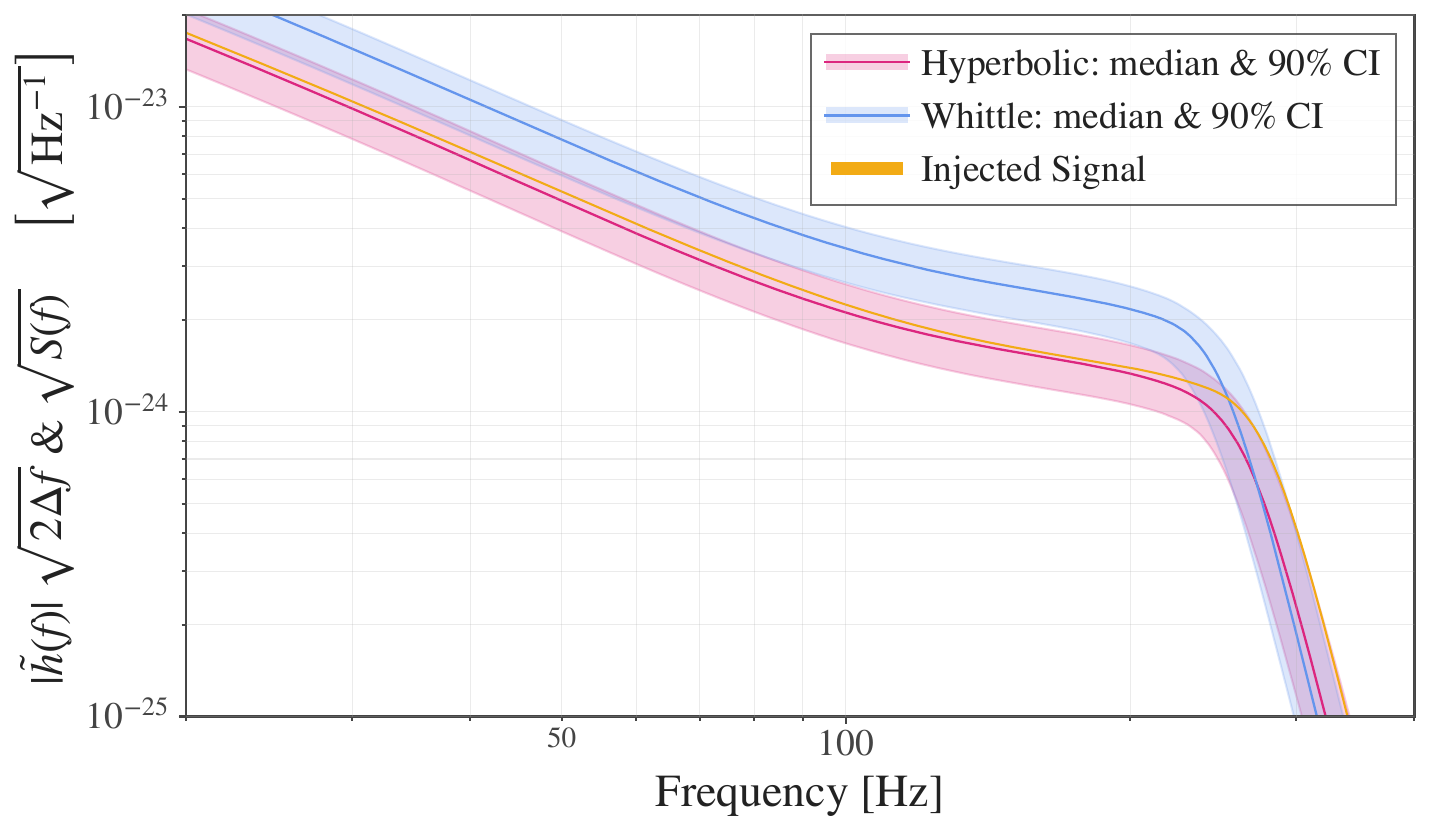}%
}

 	\caption{Similar to Fig. \ref{fig:FD_recon_long}, but for the case with the glitchy data. Only the hyperbolic likelihood provides accurate coverage of the injected signal, while the Gaussian and Whittle likelihoods fail to enclose the signal within their respective 90\% credible intervals.
}
 \label{fig:FD_recon_glitch}
\end{figure}

Fig. \ref{fig:FD_recon_glitch} illustrates the frequency domain reconstruction. In this case, only the \hyp{}  likelihood successfully recovers the injected BBH signal, with the posterior median closely tracking the true waveform
across the full frequency range and the injected signal remaining well within the
90\% C.I.

The Gaussian likelihood fails entirely in the presence of the glitch, yielding a
reconstruction that is strongly biased and does not overlap with the injected
signal. The \whtl{} likelihood provides a reconstruction that is qualitatively
closer to the true signal; however, the injected waveform lies outside the
corresponding 90\% C.I. over a substantial fraction of the frequency
band, indicating incorrect uncertainty quantification.

\begin{figure}[t]
  {%
\includegraphics[clip,width=1\columnwidth]{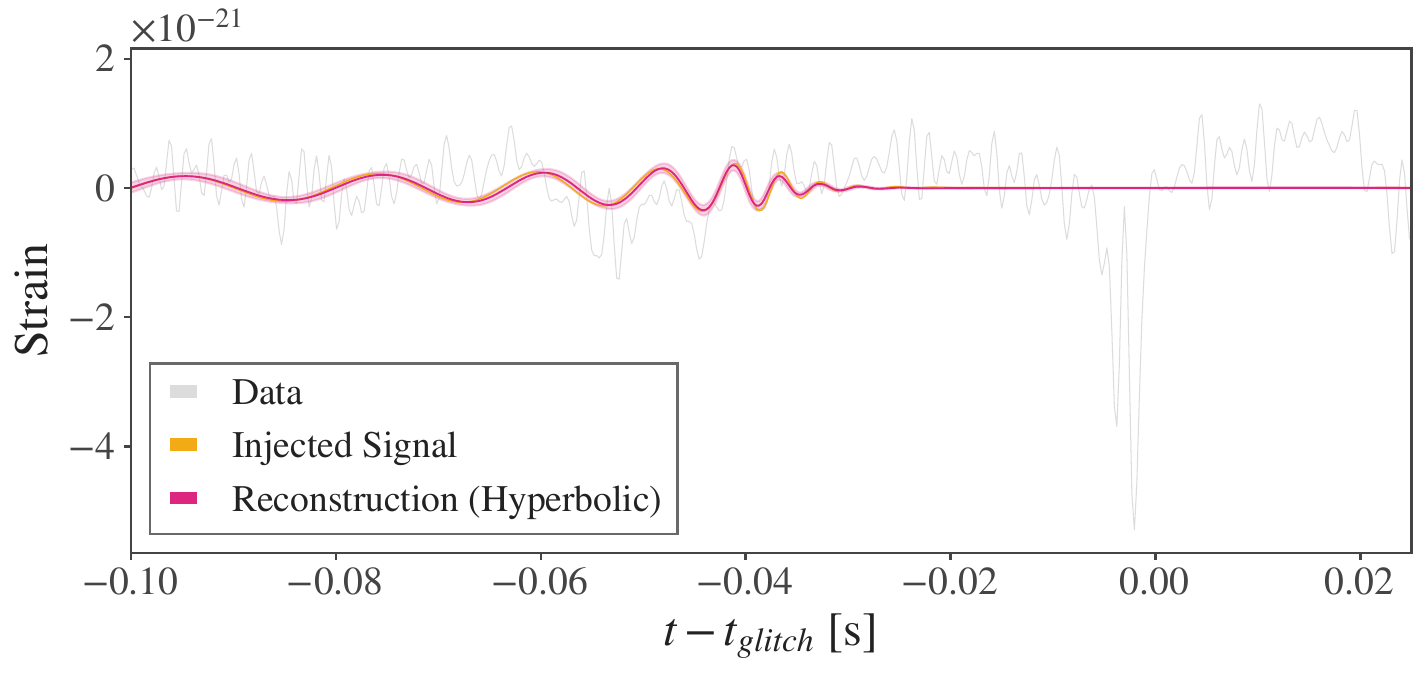}%
}
\\
{%
\includegraphics[clip,width=1\columnwidth]{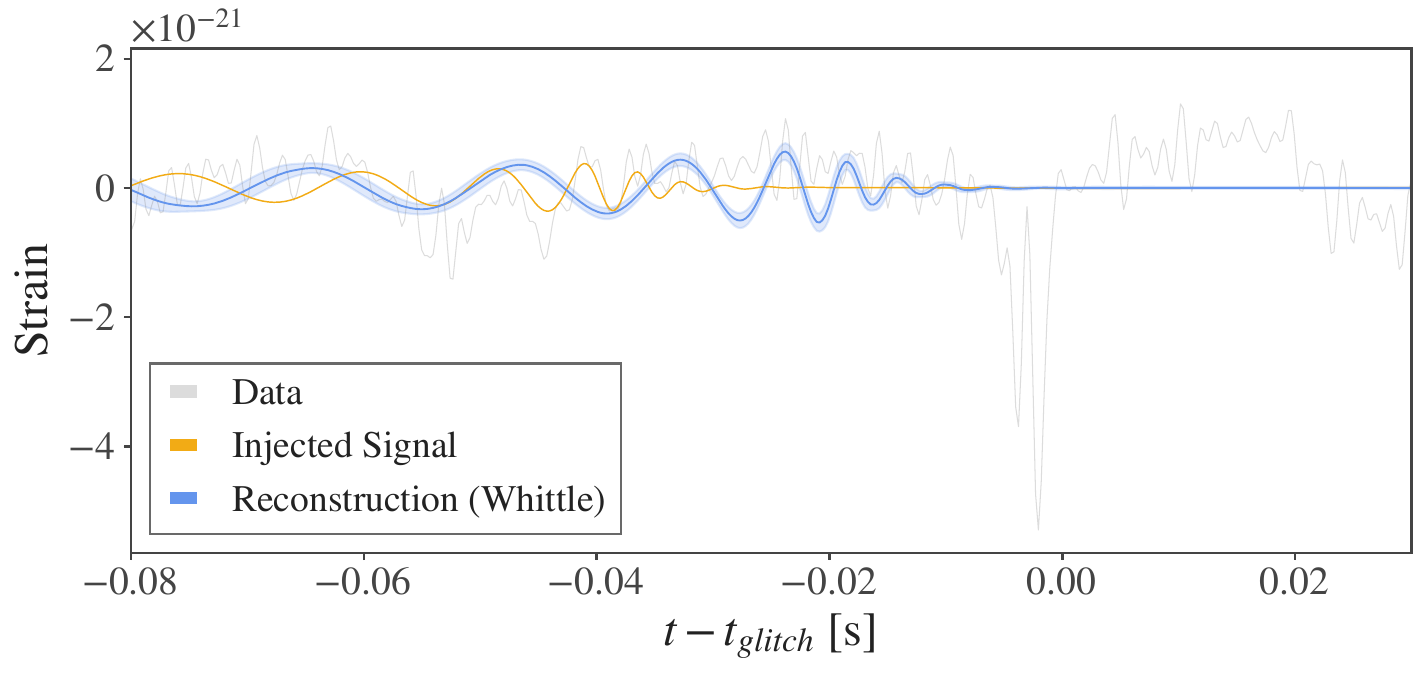}%
}

 	\caption{Time-domain reconstructions of the targeted BBH signal using the
\hyp{}  likelihood (top, pink) and the \whtl{} likelihood (bottom, blue). In each panel, the solid line shows the posterior
median and the shaded region denotes the 90\% credible interval. The injected
signal is shown in yellow, while the gray trace corresponds to the strain data.
The hyperbolic likelihood provides a reconstruction that closely follows the
injected waveform and yields reliable uncertainty coverage, whereas the Whittle
likelihood fails to recover the parameters of the injected signal within its inferred CI.
}
 \label{fig:TD_recon_glitch}
\end{figure}

Fig. \ref{fig:TD_recon_glitch} shows the time-domain reconstruction of the injected BBH signal recovered using the \hyp{}  and \whtl{} likelihoods. In the presence of the glitch, the hyperbolic likelihood yields a reconstruction which is fully encapsulated within the 90\% C.I. In contrast, the \whtl{} likelihood produces a reconstruction that is visibly
biased in phase and amplitude, with the injected signal lying outside the
corresponding C.I. over a significant portion of the time series. 

\begin{figure*}[t]
\centering
\setlength{\tabcolsep}{2pt}

\begin{subfigure}{0.31\textwidth}
\includegraphics[width=\linewidth]{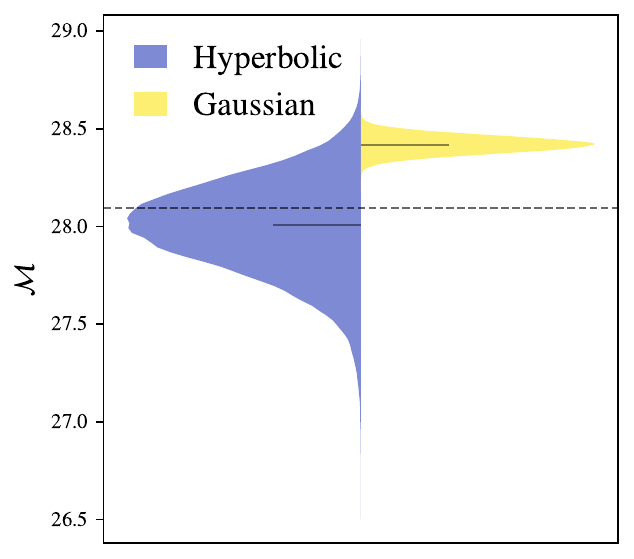}
\caption{$\mathcal{M}$~$[M_\odot]$}
\end{subfigure}\hfill
\begin{subfigure}{0.31\textwidth}
\includegraphics[width=\linewidth]{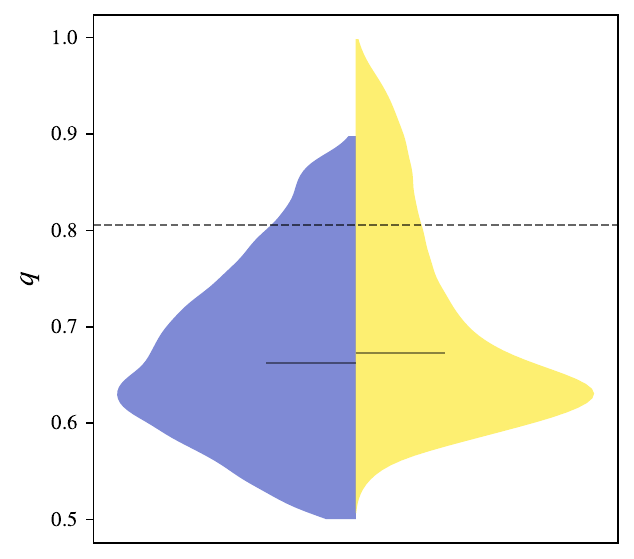}
\caption{$q$}
\end{subfigure}\hfill
\begin{subfigure}{0.31\textwidth}
\includegraphics[width=\linewidth]{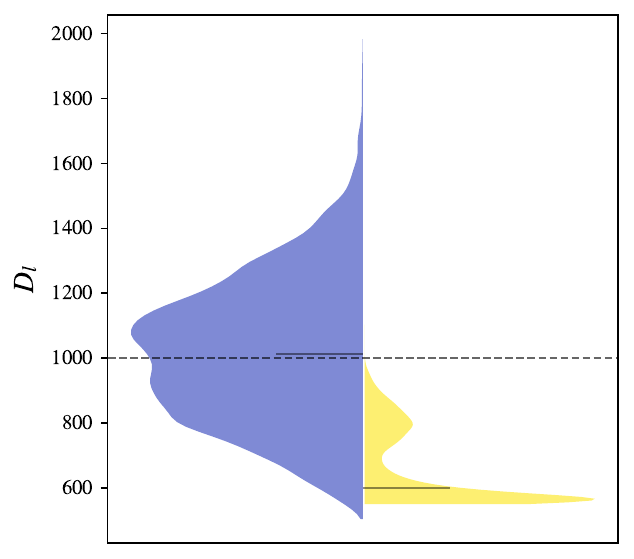}
\caption{$D_L$ [Mpc]}
\end{subfigure}

\vspace{0.3em}

\begin{subfigure}{0.31\textwidth}
\includegraphics[width=\linewidth]{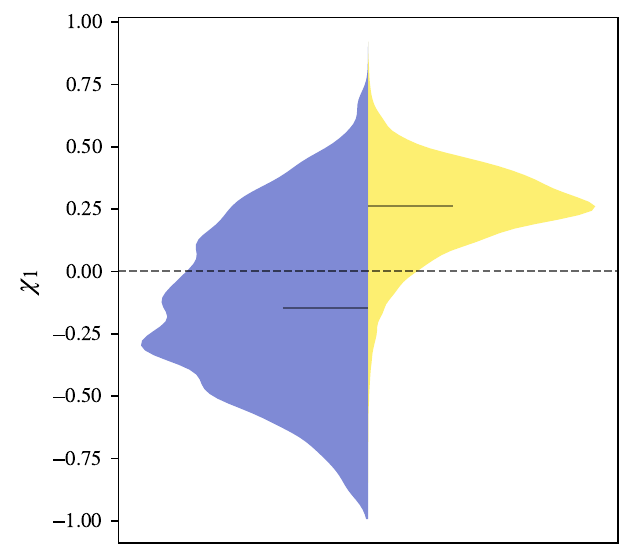}
\caption{$\chi_1$}
\end{subfigure}\hfill
\begin{subfigure}{0.31\textwidth}
\includegraphics[width=\linewidth]{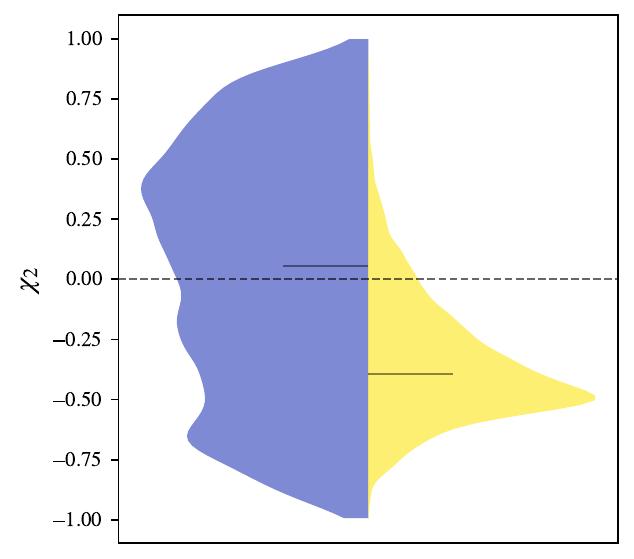}
\caption{$\chi_2$}
\end{subfigure}\hfill
\begin{subfigure}{0.31\textwidth}
\includegraphics[width=\linewidth]{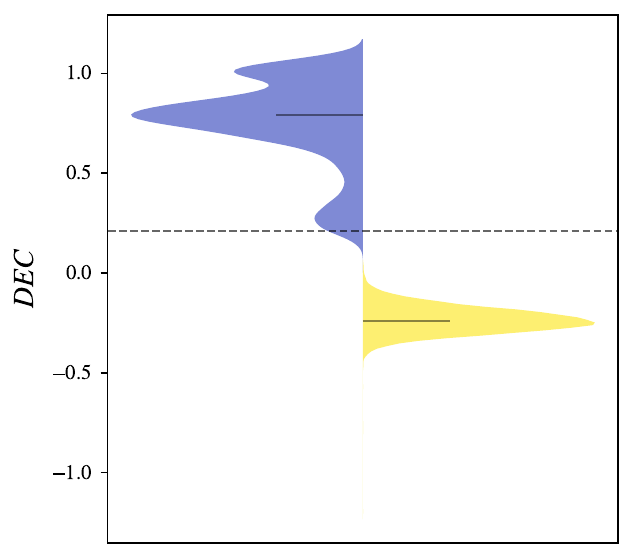}
\caption{$\mathrm{Dec}$ [rad]}
\end{subfigure}

\vspace{0.3em}

\begin{subfigure}{0.31\textwidth}
\end{subfigure}\hfill
\begin{subfigure}{0.31\textwidth}
\includegraphics[width=\linewidth]{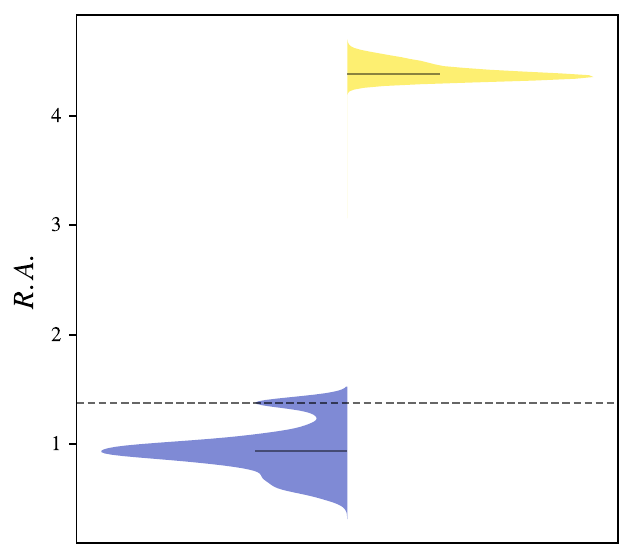}
\caption{$\mathrm{RA}$ [rad]}
\end{subfigure}\hfill
\begin{subfigure}{0.31\textwidth}
\end{subfigure}

\caption{Comparison of posterior distributions obtained with the hyperbolic and Whittle likelihoods.
Each panel shows split violin plots for key binary parameters. The hyperbolic likelihood systematically reduces bias and posterior broadening compared to the Whittle approximation.}
\label{fig:hyperwave_violin}
\end{figure*}

Fig. \ref{fig:hyperwave_violin} shows the violin plots for some key recovered parameters, useful either for population studies (chirp mass, mass ratio spins) or sky-location for follow-up strategies. The hyperbolic likelihood systematically reduces bias and
posterior broadening compared to the Whittle approximation.

\section{Conclusion \& Discussion \label{sec:conclusions}}

In our previous work \cite{Sasli:2023mxr}, we introduced a heavy-tailed likelihood based on the Generalized Hyperbolic (GH) distribution, known as hyperbolic likelihood $\Lambda_{\cal H}$. This framework was shown to provide robust parameter estimation in the presence of noise outliers, non-stationarities, and PSD model inaccuracies. Using LISA-like synthetic datasets, we demonstrated that the standard Gaussian likelihood underestimates the true noise tails, leading to over-confident posterior intervals and reduced reliability of inferred astrophysical parameters.

We subsequently extended this approach in \cite{heavy_stochastic} by incorporating a shape-agnostic PSD modeling technique based on B-splines, following a strategy similar to \cite{Baghi:2023qnq}. Within this unified heavy-tailed framework, we further analyzed stochastic GW signals in the LISA band and showed that deviations from stationarity and Gaussianity can be consistently recovered through the hyperbolic likelihood formulation and its associated spectral parameters.

Building on these developments, the present work generalizes the hyperbolic likelihood framework to the full frequency domain, providing a practical and scalable solution for robust GW parameter estimation. This extension allows us to relax the commonly adopted Gaussian noise assumption while retaining high performance under ideal conditions and significantly improving robustness in the presence of data imperfections. Using simulated LISA data with stationary Gaussian noise, we show that the \hyp{}  likelihood method performs the same as the \whtl{} likelihood under Gaussian noise conditions. Both methods provide identical parameter estimation results and demonstrate capability in reconstructing the noise PSD. 

We further apply this framework to real ground-based detector data. Our focus is on GW data featuring overlapping signals in time and frequency, as well as non-Gaussian noise transients and glitches, conditions that are expected to become ubiquitous in next-generation ground-based detectors such as Cosmic Explorer and in space-based observatories like LISA. Our investigation confirms that the Gaussian likelihood fails to recover the injected signals in these non-ideal conditions, producing biased reconstructions and severely underestimated uncertainties. The Whittle, \whtl{}, likelihood yields qualitatively improved reconstructions relative to the Gaussian case, but still suffers from incorrect coverage, when the dataset contains fast noise transients. Importantly, the hyperbolic, \hyp{} , likelihood achieves this robustness without requiring an explicit parametric model for the noise PSD. Instead, deviations from Gaussianity are captured directly through the inferred hyperbolic parameters, providing a more robust PE. This feature is particularly relevant for future detectors, where non-stationary noise, calibration uncertainties, and signal confusion are expected to be the norm rather than the exception. A corner plot comparing the \hyp{}  and \whtl{} likelihood is given in Appendix B (\ref{fig:LIGO_posterios_glitch}).

As an alternative to modifying the likelihood function, one could transform the data in order to Gaussianize them prior to analysis. An extreme case of this methodology involves the complete excision of non-Gaussian segments, such as data containing noise artifacts like glitches. More generally, it involves applying a mathematical transformation to the data before they are used, effectively enforcing a Gaussian distribution. Such data transformation strategies have been successfully applied in other domains, such as the vetting of exoplanet transits (e.g., see \cite{stab1178}, and Appendix F of \cite{robnik2026exoplanettransitsearchdetection}). However, transforming the data in a pre-processing step assumes a fixed transformation, thereby ignoring any uncertainty regarding what constitutes the optimal adjustment. This is conceptually akin to utilizing a generalized likelihood function—such as the hyperbolic likelihood we employ—but artificially fixing its parameters instead of sampling them. By jointly sampling the parameters of the hyperbolic function, our approach natively accounts for the uncertainty in the noise distribution, effectively capturing the uncertainty in the Gaussianization of the data. A rigorous quantitative comparison of these methodologies is deferred to future investigations.

To ensure computational scalability, we implemented a vectorized version of the hyperbolic likelihood evaluation. This enables parallel evaluation of the likelihood over large ensembles of parameter samples, thereby substantially reducing the computational cost of Bayesian inference in high-dimensional parameter spaces. This computational optimization facilitates the integration of the proposed framework into end-to-end inference pipelines, which will become available to the public in the near future. 

The package is currently under active development and will be made publicly available on GitHub \footnote{https://github.com/asasli/HyperWave/} upon publication of this work. The initial release will include the core likelihood implementation, PSD estimation utilities, waveform generation on both CPU and GPU architectures, and example parameter estimation workflows for real data applications. Future development plans include benchmarking against standard GW parameter estimation pipelines, extending the framework to stochastic GW searches (using LIGO data), and ensuring full compatibility with LISA and other upcoming detectors.


\begin{acknowledgments}

We wish to thank M. Katz, Q. Baghi, and Jean-Baptiste Bayle for useful comments and  fruitful discussions. We thank Xiao-Xiao Kou for useful comments on the manuscript. AS acknowledges the Bodossaki Foundation for support in the form of a PhD scholarship. AS and MWC acknowledge support from the National Science Foundation with grant numbers PHY-2308862 and PHY-2117997. NK acknowledges the funding from the European Union’s Horizon 2020 research and innovation program under the Marie Skłodowska-Curie grant agreement No 101065596. N.~S. and N.~K. acknowledge support from the Gr-PRODEX 2019 funding program (PEA 4000132310). N.~S. acknowledges funding from the H.F.R.I. Project No.~26254. In addition, this publication is part of a project that has received funding from the European Union’s Horizon Europe Research and Innovation Programme under Grant Agreement No 101131928. MK and US acknowledge funding from NSF Award Number 2311559, and from the U.S. Department of Energy, Office of Science, Office of Advanced Scientific Computing Research under Contract No. DE-AC02-05CH11231 at Lawrence Berkeley National Laboratory to enable research for Data-intensive Machine Learning and Analysis. We are grateful for the computational resources provided by the LIGO Laboratory and supported by the U.S. National Science Foundation Awards PHY-0757058 and PHY-0823459. 
Virgo is funded, through the European Gravitational Observatory (EGO), by the French Centre National de Recherche Scientifique (CNRS), the Italian Istituto Nazionale di Fisica Nucleare (INFN) and the Dutch Nikhef, with contributions by institutions from Belgium, Germany, Greece, Hungary, Ireland, Japan, Monaco, Poland, Portugal, Spain. KAGRA is supported by Ministry of Education, Culture, Sports, Science and Technology (MEXT), Japan Society for the Promotion of Science (JSPS) in Japan; National Research Foundation (NRF) and Ministry of Science and ICT (MSIT) in Korea; Academia Sinica (AS) and National Science and Technology Council (NSTC) in Taiwan.

\end{acknowledgments}

\section*{Appendices}
\subsection{Gaussian vs. Whittle in  the LISA band\label{sec:appendix A}}
In this section, we perform PE considering two scenarios. The first concerns the estimation of the MBHB parameters (\ref{tab:models}) for which we have assumed that the levels of instrumental noise are correct and fully known, while the second refers to the estimation of two additional parameters for the instrumental noise.

The methodology is similar to Sect.(\ref{sec:LISA_methodology}), but we  perform the analysis by adopting two different likelihoods; the Gaussian and the Whittle $\Lambda_{\cal W}$ (see Eq.~\ref{eq:whittle}). In addition, the noise model differs from the assumed in Sect.(\ref{sec:LISA_methodology}), here we infer the actual parameters of the SciRD noise of Eqs.(\ref{eq:s_psd_model},\ref{eq:sa_si_values}). For these parameters, we adopt uniform priors, as $\log_{10}S_a\sim\mathcal{U}\{-32, -24\}$ and $\log_{10}S_i\sim\mathcal{U}\{-25, -19\}$. 

\begin{figure*}[ht]
 	\includegraphics[width=.9\linewidth]{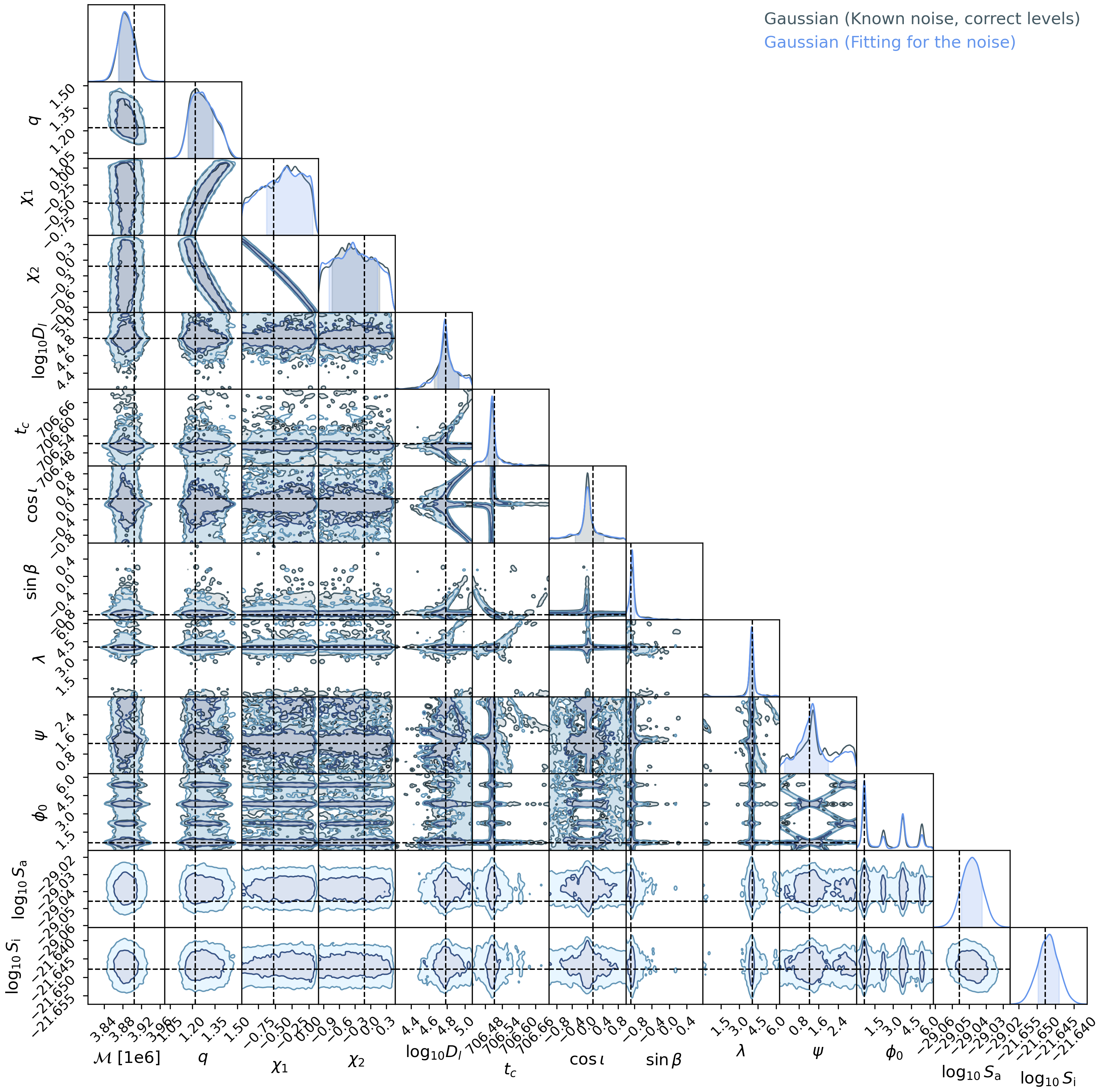}
 	\caption{Cornerplots using Gaussian likelihoods for \textbf{(a)} \textit{known noise} (grey color) and \textbf{(b)} \textit{fitting for the (unknown) noise} (light-blue color) for case in table~\ref{tab:models}).}
 \label{fig:pe_case1}
\end{figure*}

The results of the analysis are summarized in Fig.~\ref{fig:pe_case1}, and in table~\ref{tab:Summarized Results-Case 1_2}. Considering the above, we can safely conclude that the analysis for both scenarios (of known and unknown noise) yields similar estimations for the parameters. To verify this, we have also computed the Jensen-Shannon Divergence between the posteriors of each case, which are summarized in table~\ref{tab:JS Divergence}. The Jensen-Shannon divergence is a measure of how similarly distributed two-different posteriors are \cite{included_JS}. Values close to zero indicate that the distributions do not diverge, while positive values show that they diverge. We find no statistically significant differences between the posterior distributions obtained under the two likelihood assumptions.

\par
\begin{table}[ht]
\centering
\begin{tabular}{cc}
\hline
    \hline
\multirow{2}{*}{$\bm{\param}$} & {\textbf{Jensen-Shannon}}\\
        & {\textbf{Divergence}} $[10^{-3}]$ \\
		\hline
		${\cal M}$~$[M_\odot]$ & 1.24  \\
		$q$ & 0.89  \\
		$\chi_1$ & 1.01 \\
		$\chi_2$ & 1.03 \\
		$\mathrm{log}_{10}D_l$ & 9.81 \\
		$t_c$ & 17.83 \\
		$\cos\iota$ & 3.15 \\
		$\sin\beta$ & 13.13 \\
		$\lambda$ & 24.77 \\
		$\psi$ & 10.48 \\
		$\phi_0$ & 9.23 \\
  \hline
    \hline
    \end{tabular}
    \caption{Jensen-Shannon Divergence between the two likelihood types, i.e. the scenario of known (Gaussian likelihood), and unknown noise levels, Eq.(\ref{eq:whittle}), for the case in table~\ref{tab:models}. Values are reported in units of $10^{-3}$. \label{tab:JS Divergence}} 
\end{table}

\begin{table}[ht]
\begin{tabular}{l|cc}
\hline\hline
$\bm{\param_{1,2}}$ & \textbf{Known noise} & \textbf{Fitting for the noise}\\
		\hline
		${\cal M}$~$[M_\odot]$ & $\left( 388.2^{+2.3}_{-1.1} \right) \times 10^{4}$ & $\left( 388.3^{+2.1}_{-1.3} \right) \times 10^{4}$ \\
		$q$ & $1.228^{+0.107}_{-0.058}$ & $1.268^{+0.073}_{-0.095}$ \\
		$\chi_1$ & -0.3184 & $-0.33^{+0.39}_{-0.29}$ \\
		$\chi_2$ & $-0.41^{+0.59}_{-0.32}$ & $-0.34^{+0.48}_{-0.45}$ \\
		$\mathrm{log}_{10}D_l$ & $4.78^{+0.16}_{-0.12}$ & $4.779^{+0.150}_{-0.082}$ \\
		$t_c$~[hours] & $706.504^{+0.021}_{-0.024}$ & $706.504\pm 0.015$ \\
		$\cos\iota$ & $0.01^{+0.41}_{-0.33}$ & 0.0099 \\
		$\sin\beta$ & -0.855 & $-0.851^{+0.056}_{-0.069}$ \\
		$\lambda$~[rad] & $4.01^{+0.18}_{-0.21}$ & $4.01^{+0.15}_{-0.16}$ \\
		$\log_{10}S_\mathrm{a}$ & -- & $\left( -29038.1^{+5.6}_{-6.5} \right) \times 10^{-3}$ \\
		$\log_{10}S_\mathrm{i}$ & -- & $\left( -21646.6^{+2.6}_{-3.2} \right) \times 10^{-3}$ \\\hline\hline
\end{tabular}
\caption{The PE results correspond to the two different log-likelihood types for known and unknown noise levels. The left ``Known noise'' column refers to the case where the classic Gaussian likelihood is adopted, whereas the right part of the table, named ``Fitting for the noise'', refers to the analysis using the log-likelihood of Eq.(\ref{eq:whittle}).\label{tab:Summarized Results-Case 1_2} }
\end{table}

\subsection{Supplementary details of the LIGO cases\label{sec:appendix B}}

This appendix presents supplementary material supporting the multiple-injection
analyses discussed in Sect. \ref{sec:multiple sources of BBH}. We include additional corner plots and
tables of injected parameters that are not shown in the main text in order
to maintain clarity and readability. These results demonstrate the consistency
of the inferred posteriors across different injection realizations and provide
full transparency regarding the simulation setup.

For the multiple low-SNR injection study, the parameters of the additional
BBH signals were drawn randomly from astrophysically motivated prior
distributions. The injected sources span a range of masses, sky locations,
and orientations, leading to individual signal-to-noise ratios below the
typical detection threshold. The full list of injected parameters is reported
in Tables \ref{tab:model ligo} and \ref{tab:multiple injcetions} for the two different case studies of Sect. \ref{sec:multiple sources of BBH}. The injections were intentionally chosen to overlap in time and frequency
in order to stress-test the robustness of the likelihood under non-ideal noise
conditions.

\begin{figure*}[t]
\centering
\setlength{\tabcolsep}{2pt}

\begin{subfigure}{0.5\textwidth}
\includegraphics[width=\linewidth]{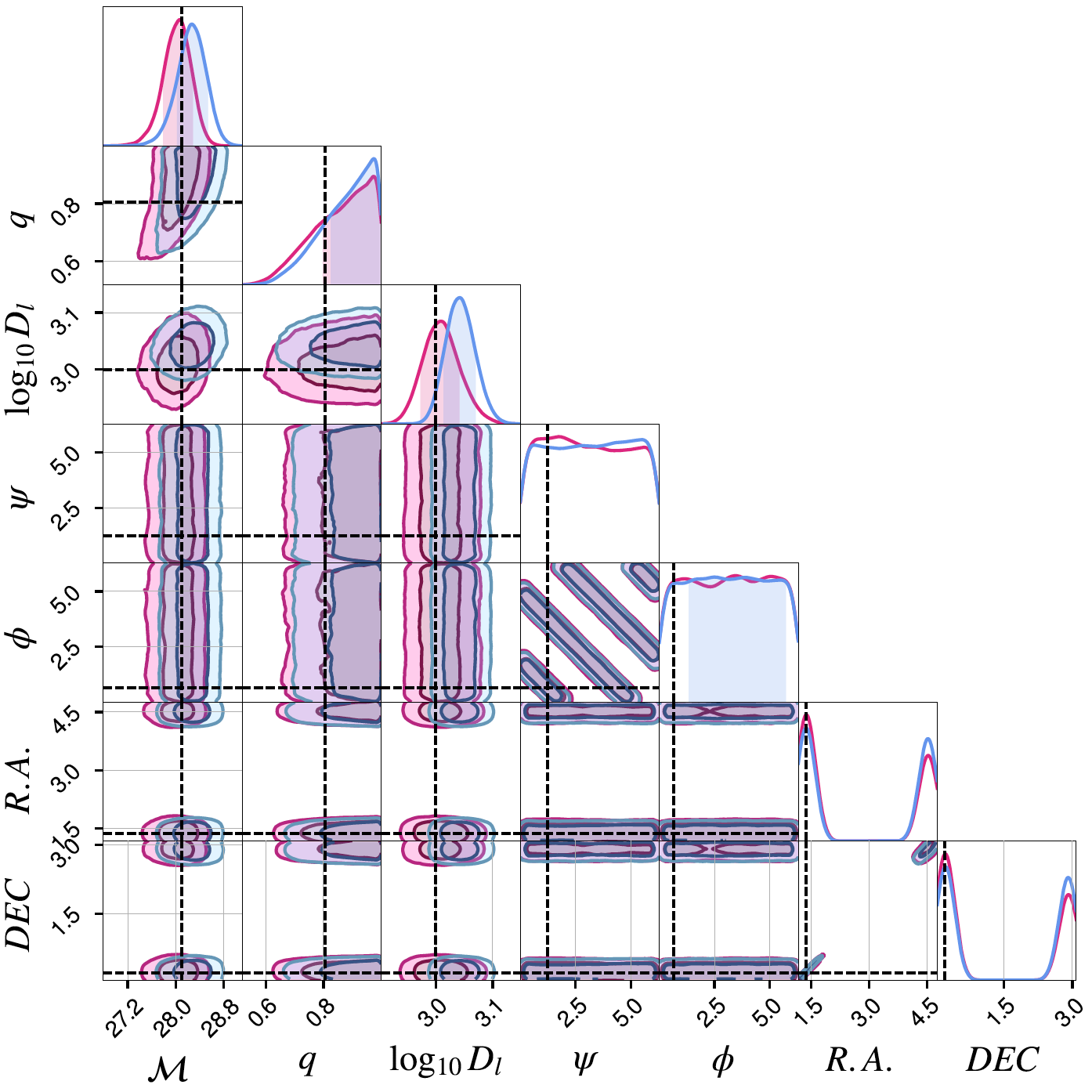}
\caption{Long-lasting BBH (see Table~\ref{tab:model_long_ligo}).}
\end{subfigure}\hfill
\begin{subfigure}{0.5\textwidth}
\includegraphics[width=0.99\linewidth]{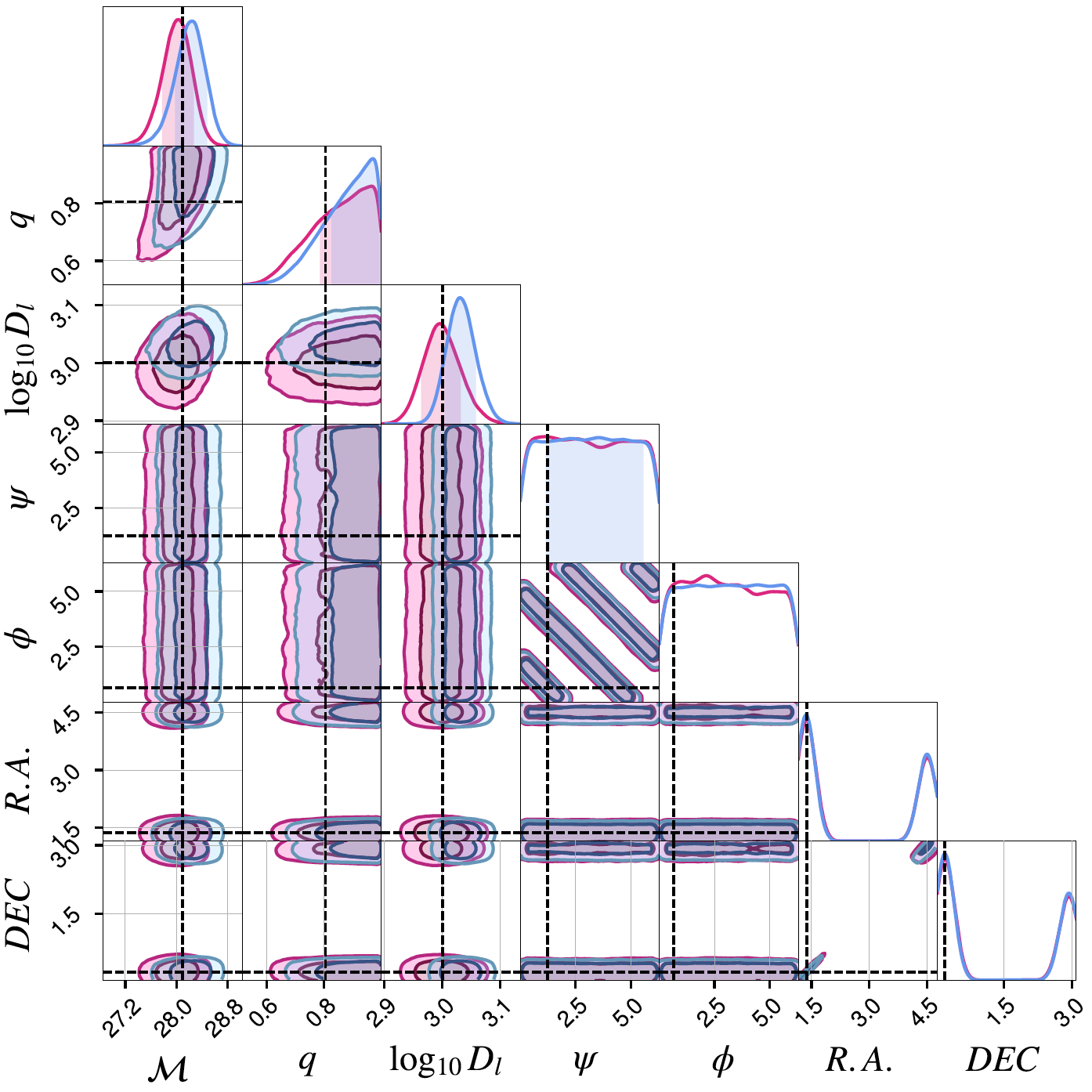}
\caption{Multiple injections of BBHs (see Table~\ref{tab:multiple injcetions}).}
\end{subfigure}\hfill

\caption{Corner plot using the \whtl{} (blue) and the \hyp{} (pink) likelihoods. The PE refers to the targeted BBH (see Table~\ref{tab:model ligo}) when we simultaneously inject (a) a long-lasting BBH and (b) multiple low-SNR BBHs.}
\label{fig:LIGO_posterios_long_and_multi_bbh}
\end{figure*}

Fig.~\ref{fig:LIGO_posterios_long_and_multi_bbh} shows the posterior distributions
for the targeted BBH parameters obtained in the presence of multiple
overlapping injections. The \hyp{}  and \whtl{}  likelihoods are shown for comparison. While the overall structure of the posteriors is almost similar, the \hyp{}  likelihood exhibits increased robustness against contamination
from unresolved signals, whereas the \whtl{} likelihood displays a mild tendency toward biased inference.

\begin{figure}[t]
  \subfloat[Hyperbolic (pink) and Gaussian (green) likelihood]{%
\includegraphics[clip,width=0.95\columnwidth]{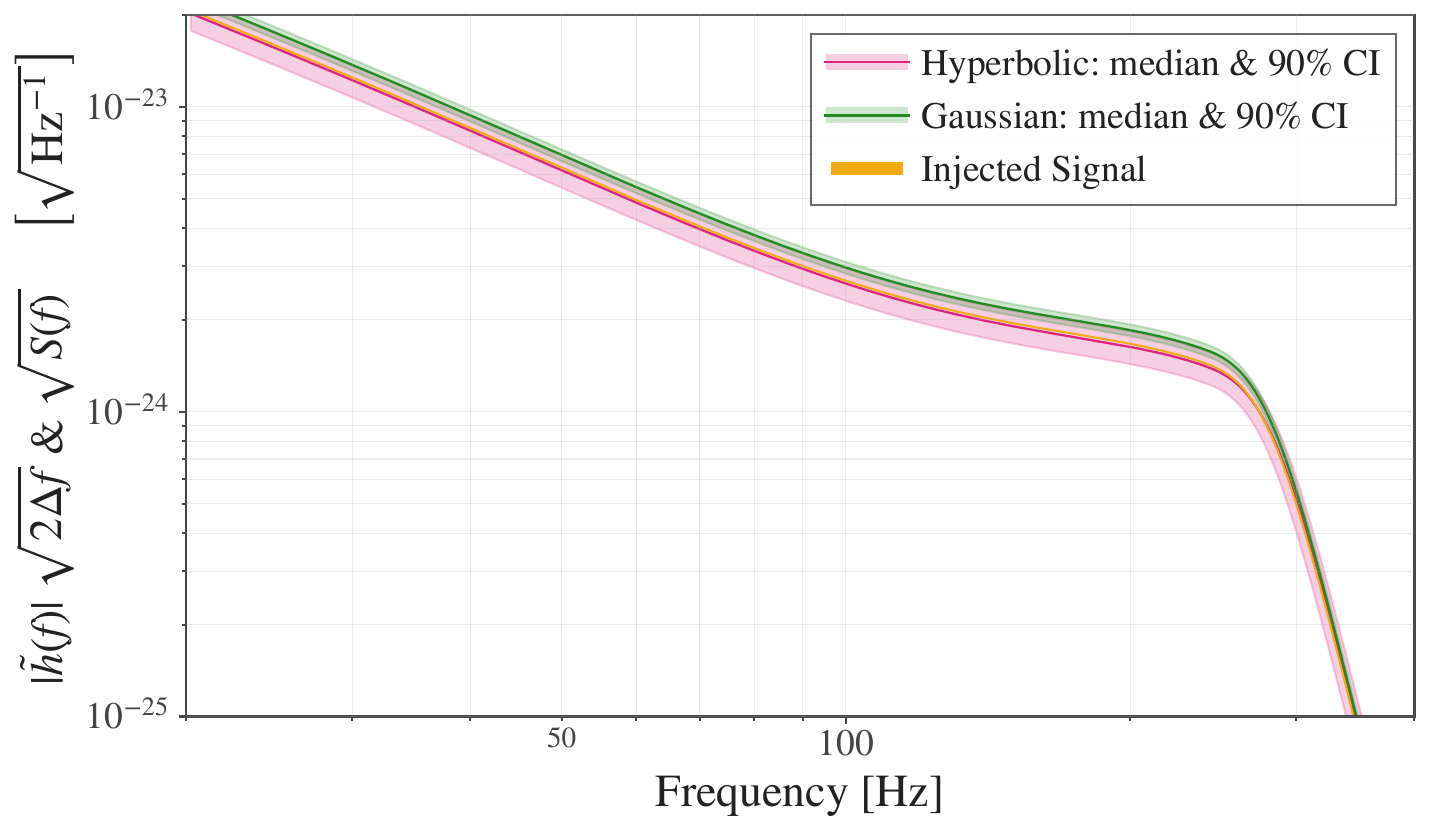}%
}

    \subfloat[Whittle likelihood]{%
\includegraphics[clip,width=0.95\columnwidth]{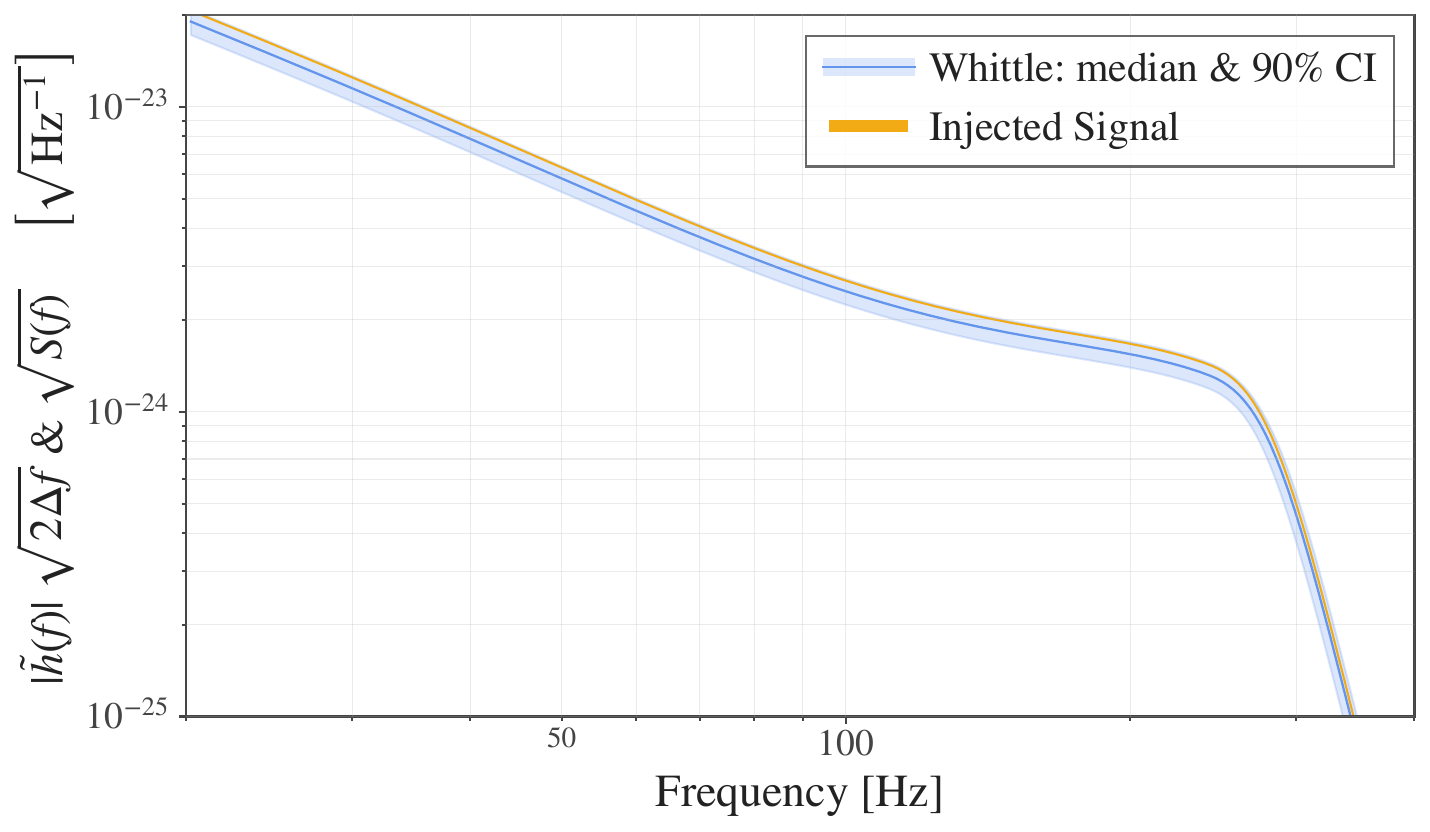}%
}

 	\caption{Reconstruction (median and 90\% C.I. reported) in the frequency domain of the targeted BBH merger in H1 detector. This case refers to the data with the long-lasting BBH injected. The reconstrucion presented concerns the PE by adopting the {\it (a)} \hyp{} and Gaussian and {\it (b)} the \whtl{} likelihood.
}
 \label{fig:FD_recon_long}
\end{figure}

\begin{figure}[t]
  \subfloat[Hyperbolic (pink) and Gaussian (green) likelihood]{%
\includegraphics[clip,width=0.95\columnwidth]{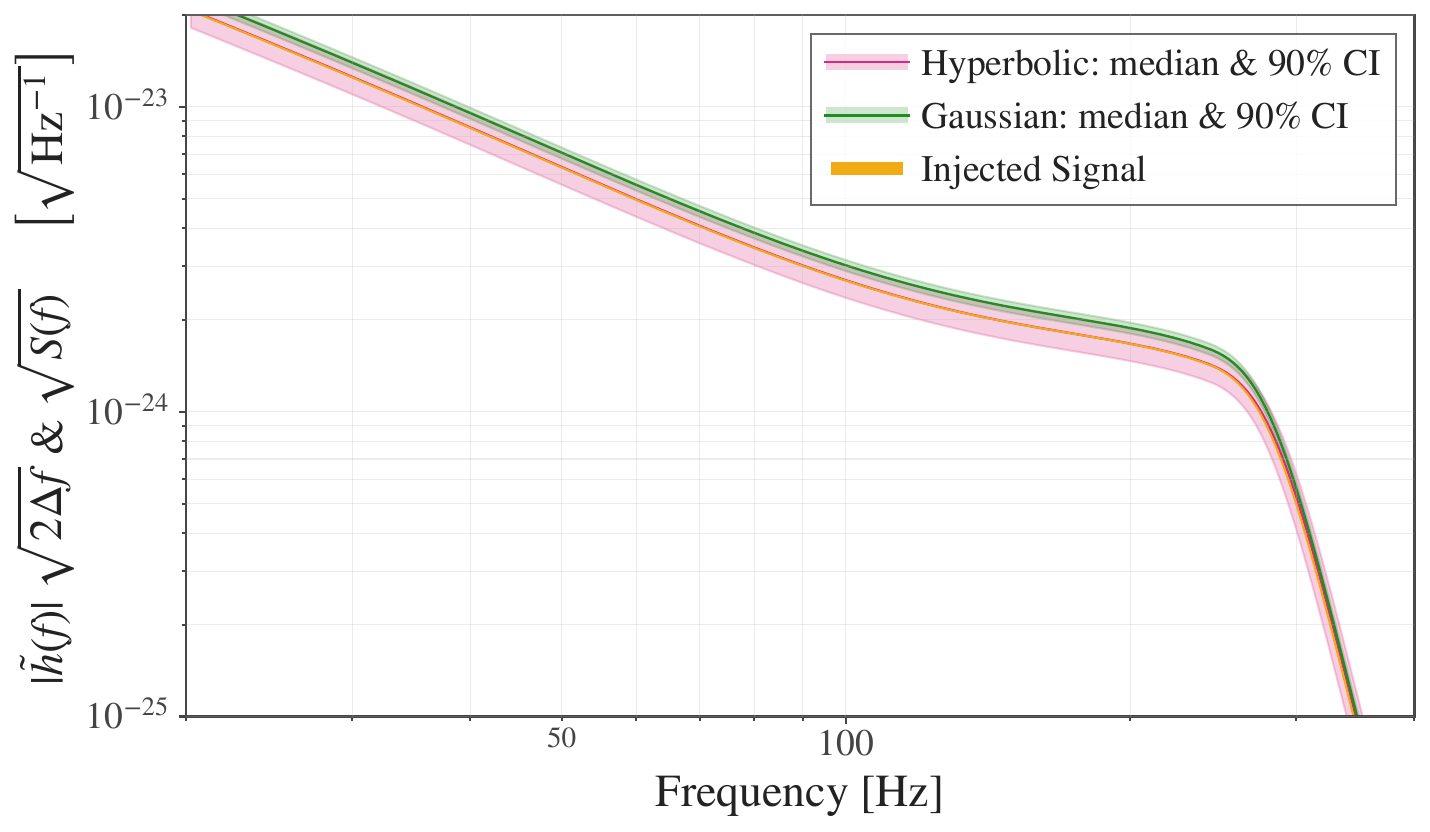}%
}

    \subfloat[Whittle likelihood]{%
\includegraphics[clip,width=0.95\columnwidth]{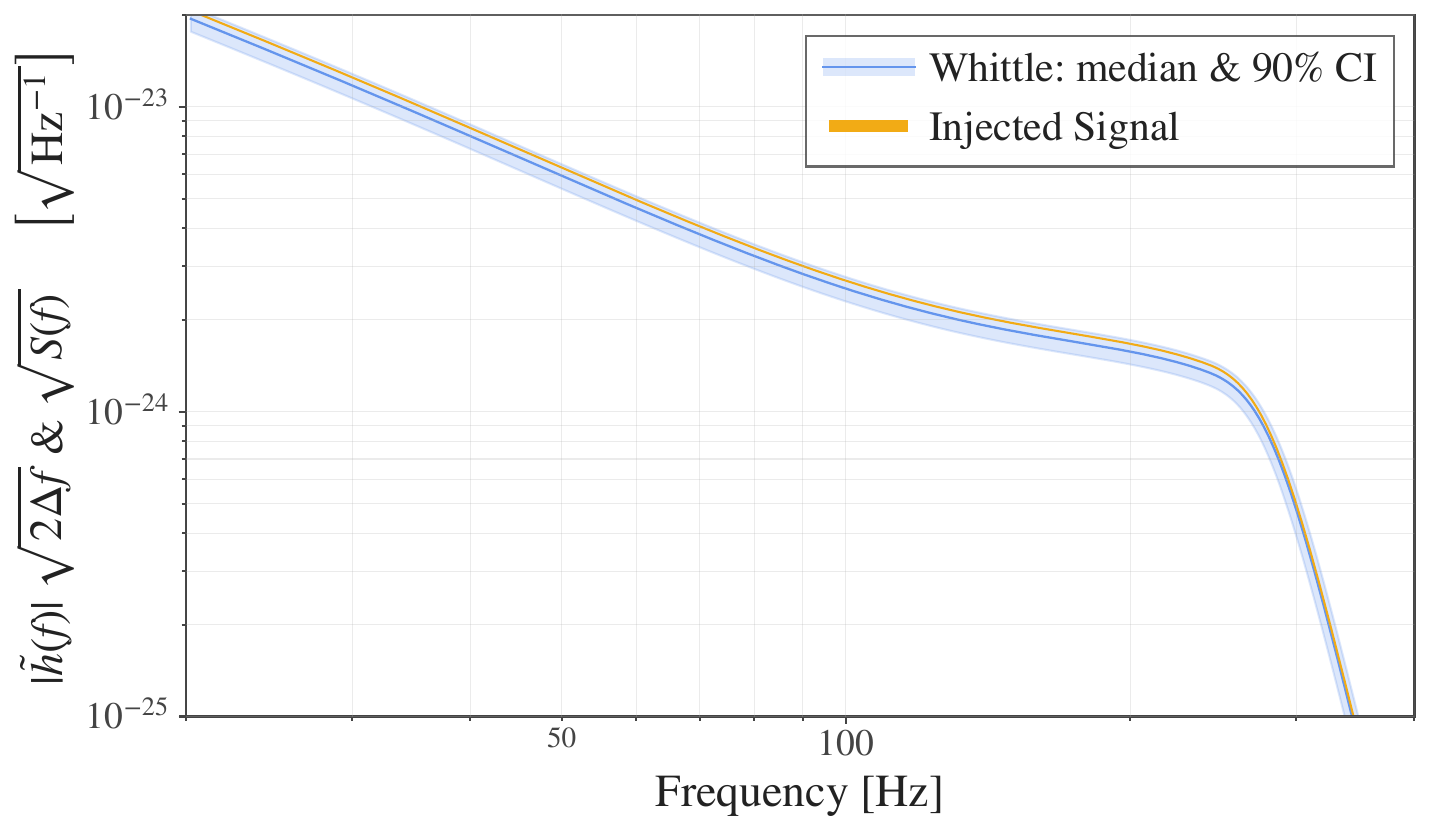}%
}

 	\caption{Similar to Fig. \ref{fig:FD_recon_long}, but for the case with the seven low-SNR BBH injections.
}
 \label{fig:FD_recon_multi}
\end{figure}

Fig. \ref{fig:FD_recon_long} and~\ref{fig:FD_recon_multi} demonstrate that the \hyp{}  likelihood provides a stable and
accurate reconstruction, with the median reconstructed almost perfectly overlapping with the injected signal. The Gaussian likelihood (which is the one that most GW pipelines use) fails to recover the parameters of the injected signal, with significant
portions of the injected waveform lying outside the inferred credible region. The \whtl{} likelihood captures the injected signal within the 90\% credible interval.

Despite the substantially different configurations of overlapping sub-threshold BBH signals, the resulting posterior distributions are strikingly similar in both experiments. This robustness indicates that the inference is primarily driven by the common high-SNR BBH signal and the common (real) noise, while the contribution of unresolved sub-threshold signals is effectively incorporated into the noise model. 

\begin{table}[ht]
\caption{Parameters of the long-lasting BBH injection in LIGO data.}
\label{tab:model_long_ligo}
\begin{ruledtabular}
\begin{tabular}{lc}
\textbf{Parameter} & \textbf{Value} \\
\hline
$\mrm{SNR_{L1}}$ & 3.42 \\
$\mrm{SNR_{H1}}$ & 3.39 \\
Mass 1, $m_1~[M_\odot]$ & 15  \\ 
Mass 2, $m_2~[M_\odot]$ & 8   \\
Distance, $D$ [Mpc] & 2000  \\ 
Right Ascension (RA) [rad] & 2.75 \\
Declination (Dec) [rad] & 1.79  \\
Polarization, $\psi$ [rad] & 0.60 \\
Phase at coalescence, $\phi_0$ [rad] & 3.91 \\
Inclination angle [rad] & 0.40  \\
Primary object tilt [rad] & 0 \\
Secondary object tilt [rad] & 0 \\
Primary dimensionless spin magnitude & 0 \\
Secondary dimensionless spin magnitude & 0 \\
Relative spin azimuthal angle, $\phi_{12}$ [rad] & 0 \\
Spin phase angle, $\phi_{jl}$ [rad] & 0 \\
\end{tabular}
\end{ruledtabular}
\end{table}

\begin{table*}[t]
\centering
\caption{Parameters of the low SNR BBH injections.\label{tab:multiple injcetions}}
\begin{ruledtabular}
    \begin{tabular}{c@{\hskip 0.2in}c@{\hskip 0.2in}c@{\hskip 0.2in}c@{\hskip 0.2in}c@{\hskip 0.2in}c@{\hskip 0.2in}c@{\hskip 0.2in}c@{\hskip 0.2in}c@{\hskip 0.2in}c}
$\bm{\#}$ &
$\bm{\mrm{SNR_{L1}}}$ &
$\bm{\mrm{SNR_{H1}}}$ &
$\bm{\mathcal{M}}$~[$M_\odot$] &
$\bm{\mrm{q}}$ &
$\mathbf{\log_{10} D_{l}}$\, [Mpc] &
$\bm{\psi}$ [rad] &
$\bm{\phi_0}$ [rad] &
\textbf{ra} [rad] &
\textbf{dec} [rad] \\
      \hline
    1 & 4.4 & 2.6 & 17.37 & 0.91 & 3.46 & 1.93 & 1.99 & 4.43 & -2.84  \\ 
    2 & 5.4 & 4.4 & 18.61 & 0.86 & 3.46 & 0.24 & 3.57 & 1.37 & 1.10 \\
    3 & 1.9 & 3.4 & 18.04 & 0.87 & 3.46 & 1.16 & 5.46 & 5.81 & 0.60  \\
    4 & 3.1 & 2.6 & 17.39 & 0.70 & 3.45 & 2.93 & 2.74 & 2.78 & 0.21 \\
    5 & 1.3 & 2.4 & 17.76 & 0.73 & 3.47 & 2.05 & 5.04 & 5.71 & -2.87  \\
   6 & 4.8 & 4.2 & 17.26 & 0.86 & 3.47 & 1.25 & 0.90 & 0.38 & 0.39 \\
    7 & 2.6 & 2.4 & 17.34 & 0.87 & 3.46 & 2.48 & 4.42 & 1.16 & -1.07 \\ 
    \end{tabular}
\end{ruledtabular}
\end{table*}

\begin{figure*}[t]
 \includegraphics[width=\linewidth]{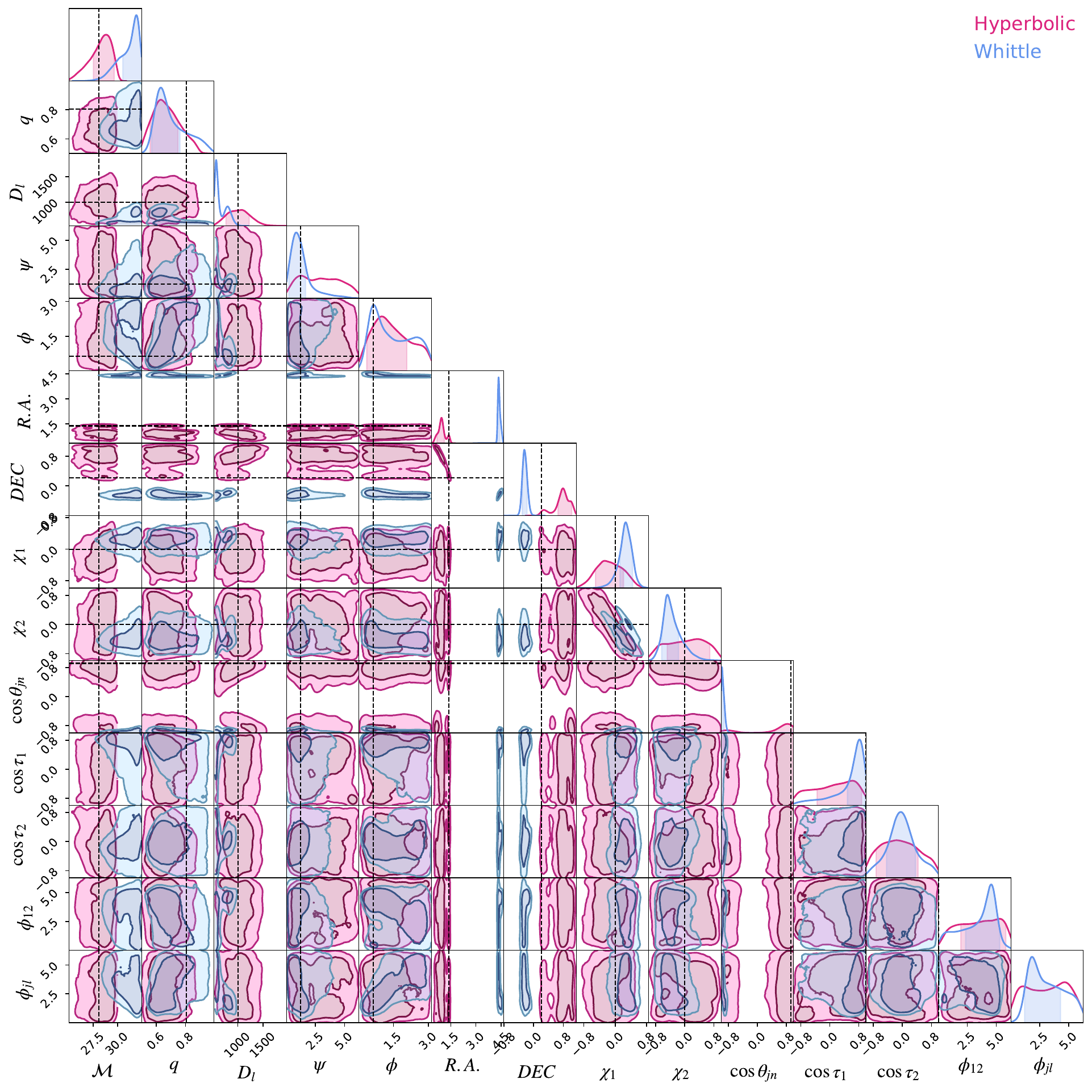}
 	\caption{Corner plot using the \whtl{} (blue) and the \hyp{}  (pink) likelihood. This case corresponds to the case of the dataset with the ``blip'' glitch. The parameters of the injected BBH are given in Table~\ref{tab:model ligo}.}
 \label{fig:LIGO_posterios_glitch}
\end{figure*}

These results suggest that the PSD correction inferred through the Whittle and Hyperbolic likelihoods effectively regularizes the inference problem by absorbing the contribution of unresolved BBH signals into an effective noise floor. As a consequence, the identifiability of the high-SNR BBH parameters is preserved even in the presence of complex signal overlap. Moreover, the improved stability of the \hyp{}  formulation highlights its enhanced ability to capture residual non-Gaussian features and PSD mismodeling effects, which become particularly relevant in scenarios dominated by overlapping sub-threshold signals. In regimes characterized by stronger non-Gaussian noise features, the \whtl{} likelihood tends to produce increasingly biased posterior estimates, indicating a breakdown of the Gaussian approximation underlying the Whittle formulation (as shown in Figures \ref{fig:FD_recon_glitch}, \ref{fig:TD_recon_glitch}, \ref{fig:hyperwave_violin} and \ref{fig:LIGO_posterios_glitch}). 

Fig.~\ref{fig:LIGO_posterios_glitch} shows the posterior distributions
for the targeted BBH parameters obtained in the presence of the blip ``glitch''. In this glitch-dominated regime, the Whittle likelihood—which assumes Gaussian stationary noise—struggles to disentangle signal and noise contributions, leading to biased results, especially for the $\mathcal{M}$ and distance. Importantly, the Hyperbolic posteriors remain consistent with the injected parameter values across most dimensions. Overall (and given the reconstruction plots \ref{fig:FD_recon_glitch} and \ref{fig:TD_recon_glitch}), the Hyperbolic likelihood provides a more reliable and interpretable posterior structure in the presence of glitches, making it particularly suitable for real detector data.

\end{document}